\documentclass[12pt,a4paper]{iopart}

\usepackage{iopams}
\usepackage{setstack}
\usepackage{graphicx}
\usepackage{pict2e}
\usepackage{color}
\usepackage{hyperref}

\renewcommand{\P}{\mathbb{P}}
\newcommand{\Li}{\mathrm{Li}}

\renewcommand{\Im}{\mathrm{Im}}
\newcommand{\sign}{\mathrm{sgn}}
\newcommand{\Z}{\mathbb{Z}}
\newcommand{\C}{\mathbb{C}}
\newcommand{\Ch}{\widehat{\mathbb{C}}}
\newcommand{\D}{\mathbb{D}}
\newcommand{\R}{\mathcal{R}}
\newcommand{\RN}{\R_{N}}
\newcommand{\RNp}{\mathring{\R}_{N}}
\renewcommand{\[}{[\![}
\renewcommand{\]}{]\!]}

\newcommand{\Xint}[1]{
	\mathchoice
		{\,\XXint{\displaystyle}{\textstyle}{#1}}%
		{\,\XXint{\textstyle}{\scriptstyle}{#1}}%
		{\XXint{\scriptstyle}{\scriptscriptstyle}{#1}}%
		{\XXint{\scriptscriptstyle}{\scriptscriptstyle}{#1}}%
	\!\int
}
\newcommand{\XXint}[3]{{\setbox0=\hbox{$#1{#2#3}{\int}$}\vcenter{\hbox{$#2#3$}}\kern-0.5\wd0}}
\newcommand{\rint}{\Xint-}

\newcommand{\keywords}[1]{\noindent\textbf{Keywords:} #1}
\newcommand{\tfrac}[2]{\mbox{\small$\frac{#1}{#2}$}}
\renewcommand{\text}[1]{\mathrm{#1}}

\begin{document}

\title{Riemann surface for TASEP with periodic boundaries}
%\titlerunning{}
\author{Sylvain Prolhac}
\address{Laboratoire de Physique Th\'eorique, IRSAMC, UPS, Universit\'e de Toulouse, France}

%\maketitle

\begin{abstract}
The Bethe ansatz solution of periodic TASEP is formulated in terms of a ramified covering from a Riemann surface to the sphere. The joint probability distribution of height fluctuations at $n$ distinct times has in particular a relatively simple expression as a function of $n$ variables on the Riemann surface built from exponentials of Abelian integrals, traced over the ramified covering and integrated on $n$ nested contours in the complex plane.\\


%\vspace{10mm}
%\today
\end{abstract}

%%%%%%%%%%%%%%%%%
%%             %%
%%  Section I  %%
%%             %%
%%%%%%%%%%%%%%%%%
\begin{section}{Introduction}
The totally asymmetric simple exclusion process (TASEP) \cite{D1998.1,S2001.1,GM2006.1,M2011.1} is a Markov process featuring hard-core particles hopping asymmetrically between neighbouring sites of a lattice. In the one-dimensional model with periodic boundary conditions studied in this paper, the particles hop with constant rate $1$ from any site $i$ to the next site $i+1$.

At large scales, TASEP belongs to KPZ universality \cite{KPZ1986.1,HHZ1995.1,KK2010.1,C2011.1,QS2015.1,S2016.2,S2019.1}. More precisely, calling $L$ the number of lattice sites and $N$ the number of particles, the statistics of the height function of TASEP at fixed density $\rho=N/L$ converges at large $L$ on the time scale $t\sim L^{3/2}$ to that of the KPZ fixed point in finite volume, describing how Tracy-Widom distributions and Airy processes characteristic of the process on the infinite line \cite{J2000.1,BR2000.1,S2005.1,FS2006.1,FF2011.1,CQR2011.1,FS2016.1,MQR2017.1,DNLDT2017.1,QR2019.1,JR2019.1} relax \cite{MSS2012.1,P2016.1,BL2018.1,L2018.1} to a Brownian stationary state with non-Gaussian large deviations \cite{DL1998.1,BD2000.1,PPH2003.1,GLMV2012.1,MP2018.1}. In this paper, we revisit height fluctuations of TASEP with periodic boundaries from the point of view of algebraic geometry.

TASEP is an integrable model, and the eigenfunctions of the time evolution operator are obtained by the Bethe ansatz. The Bethe equations of TASEP have a peculiar mean field structure, which we relate in this paper to the existence of a covering map $\pi_{N}$ from a compact Riemann surface $\RN$ to the Riemann sphere $\Ch$, underlying the integrability of the model. The joint probability distribution at $n$ distinct times of the TASEP height is in particular expressed in (\ref{P[H>=U] exp}) as a function of $n$ variables on $\RN$ built from exponentials of Abelian integrals, traced over $\pi_{N}$ and integrated on $n$ nested contours on $\C$. This is our main result, valid under the hypothesis that $L$ and $N$ are co-prime to avoid technicalities arising when the Riemann surface $\RN$ has several connected components, and derived using standards tools from quantum integrability. The large $L$ asymptotics to the KPZ regime then follows rather easily from (\ref{P[H>=U] exp}), as the joint probability of the height at the KPZ fixed point may be expressed in a similar way, with $\RN$ converging in some sense to the non-compact Riemann surface $\R_{\text{KPZ}}$ build from half-integer polylogarithms \cite{P2020.1}.

Equivalent expressions for the joint probability of the height of TASEP and the corresponding large $L$ limit in the KPZ regime were obtained earlier by Baik and Liu \cite{BL2019.1,BL2019.2} using the more rigorous propagator approach. On the other hand, the expansion over Bethe eigenstates used in this paper allows us to use standard results from quantum integrability to make several parts of the derivation easier. Additionally, our interpretation of the contribution of each eigenstate as an integral over the Riemann surface $\RN$ leads to much simpler expressions with clear analytic structure. We expect that our Riemann surface approach will be useful for future extensions, in particular to other kinds of boundary conditions.

The paper is organized as follows. In section~\ref{section Bethe ansatz}, we recall Bethe ansatz formulas for the eigenstates of TASEP and their scalar products. In section~\ref{section Riemann surfaces}, we introduce the Riemann surface $\RN$ and the covering map $\pi_{N}$ from $\RN$ to the sphere, and study some of their properties. Finally, in section~\ref{section Height fluctuations}, we study height fluctuations for TASEP and state our main result (\ref{P[H>=U] exp}). Some technical calculations are gathered in appendix.
\end{section}

%%%%%%%%%%%%%%%%%%
%%              %%
%%  Section II  %%
%%              %%
%%%%%%%%%%%%%%%%%%
\begin{section}{Bethe ansatz for TASEP}
\label{section Bethe ansatz}
In this section, we recall known results about the Bethe ansatz integrability of TASEP for a system with $N$ particles on $L$ sites and periodic boundary conditions. Configurations $\mathcal{C}$ of the system can be specified either by the occupation numbers $n_{1},\ldots,n_{L}\in\{0,1\}$, with $n_{i}=0$ (respectively $n_{i}=1$) corresponding to an empty site (resp. an occupied site), or by the positions $x_{j}$ of the particles on the lattice, $1\leq x_{1}<\ldots<x_{N}\leq L$. The set of all configurations $\Omega_{L,N}$ has cardinal $|\Omega_{L,N}|={{L}\choose{N}}$. We assume in the following that there is at least a particle and an empty site in the system, so that $1\leq N\leq L-1$.

\begin{subsection}{Bethe equations, eigenvalue and momentum}
The dynamics of TASEP can be described in terms of the deformed Markov matrix $M(\gamma)$ acting on configuration space by $M(\gamma)|\mathcal{C}\rangle=\sum_{\mathcal{C}'\neq\mathcal{C}}w_{\mathcal{C}'\leftarrow\mathcal{C}}\,(\rme^{\gamma}\,|\mathcal{C}'\rangle-|\mathcal{C}\rangle)$, where $w_{\mathcal{C}'\leftarrow\mathcal{C}}=1$ if there exists a site $i$ such that $\mathcal{C}'$ may be reached from $\mathcal{C}$ by moving a particle from site $i$ to $i+1$, and $w_{\mathcal{C}'\leftarrow\mathcal{C}}=0$ otherwise. When $\gamma=0$, $M(0)$ reduces to the Markov matrix of TASEP from which probabilities $P_{t}(\mathcal{C})$ that the system is in configuration $\mathcal{C}$ at time $t$ evolve. The fugacity $\gamma$ counts the current of particles across the system, and is necessary for studying the height function associated to TASEP, see section~\ref{section Height fluctuations}.

The matrix $M(\gamma)$ is related by a similarity transformation to the Hamiltonian of an XXZ spin chain with anisotropy $\Delta=\infty$ and twisted boundary conditions, and can be diagonalized exactly using Bethe ansatz. Periodicity in space implies that the momenta $q_{j}$ of quasi-particles are quantized, and the quantities $y_{j}=1-\rme^{\rmi q_{j}-\gamma}$ must be solution of the Bethe equations
\begin{equation}
\label{Bethe equations}
\rme^{L\gamma}(1-y_{j})^{L}=(-1)^{N-1}\prod_{k=1}^{N}\frac{y_{j}}{y_{k}}\;.
\end{equation}
The Bethe equations of TASEP have a mean field nature, with $y_{j}$ being coupled to the other $y_{k}$ only through the symmetric function $\prod_{k=1}^{N}y_{k}$. This is a consequence of the anisotropy $\Delta=\infty$ in the corresponding XXZ spin chain, or equivalently of the mapping to a five vertex model. This observation, which was crucial in many earlier works \cite{GS1992.1,DL1998.1} on the model, is the key point leading to the Riemann surface $\RN$ in section~\ref{section Riemann surfaces}.

According to coordinate Bethe ansatz, eigenvectors of $M(\gamma)$ are given in terms of the Bethe roots $y_{j}$ by \cite{D1998.1}
\begin{eqnarray}
\label{psi r}
&& \langle x_{1},\ldots,x_{N}|\psi_{\vec{y}}(\gamma)\rangle=\det(y_{j}^{-k}(1-y_{j})^{x_{k}}\rme^{\gamma x_{k}})_{j,k\in\[1,N\]}\\
\label{psi l}
&& \langle\psi_{\vec{y}}(\gamma)|x_{1},\ldots,x_{N}\rangle=\det(y_{j}^{k}(1-y_{j})^{-x_{k}}\rme^{-\gamma x_{k}})_{j,k\in\[1,N\]}\;.
\end{eqnarray}
Since $M(\gamma)$ is not a symmetric matrix, the left and right eigenvectors are not transpose of each other. In our notations $|\psi_{\vec{y}}(\gamma)\rangle$ and $\langle\psi_{\vec{y}}(\gamma)|$ above, the variable $\gamma$ refers only to the explicit parameter $\gamma$ in the determinants (\ref{psi r}), (\ref{psi l}), and not to the fact that $\vec{y}=(y_{1},\ldots,y_{N})$ must be solution of the Bethe equations (\ref{Bethe equations}) with the same parameter $\gamma$ in order for the Bethe vectors (\ref{psi r}), (\ref{psi l}) to be eigenvectors of $M(\gamma)$. In section~\ref{section scalar products} below, we also consider Bethe vectors with parameters $y_{j}$ \emph{not} solution of Bethe equations, which are needed in section~\ref{section Height fluctuations} for height fluctuations.

Given a solution $\vec{y}$ of the Bethe equation (\ref{Bethe equations}), the eigenvalue of $M(\gamma)$ corresponding to the left and right eigenvectors $\langle\psi_{\vec{y}}(\gamma)|$ and $|\psi_{\vec{y}}(\gamma)\rangle$ is equal to
\begin{equation}
\label{E[y]}
E(\gamma)=\sum_{j=1}^{N}\frac{y_{j}}{1-y_{j}}\;.
\end{equation}
Additionally, the matrix $M(\gamma)$ commutes with the translation operator $T$ defined by $T|x_{1},\ldots,x_{N}\rangle=|x_{1}-1,\ldots,x_{N}-1\rangle$, and the Bethe vectors (\ref{psi r}), (\ref{psi l}) are eigenvectors of $T$ with eigenvalue
\begin{equation}
\label{P[y]}
\rme^{\rmi P/L}=\rme^{N\gamma}\prod_{j=1}^{N}(1-y_{j})\;,
\end{equation}
with $P\in2\pi\Z$ the momentum of the eigenstate, defined modulo $2\pi L$.
\end{subsection}

\begin{subsection}{Scalar products of Bethe states}
\label{section scalar products}
The expressions (\ref{psi r}), (\ref{psi l}) for the eigenvectors are known as symmetric Grothendieck polynomials in the Bethe roots \cite{MS2013.1}. The Cauchy identity for the off-shell / off-shell scalar product, between Bethe vectors $\langle\psi_{\vec{w}}(\gamma)|$ and $|\psi_{\vec{y}}(\gamma)\rangle$ with arbitrary parameters $w_{j}$, $y_{j}$ not necessarily solution of the Bethe equations is \cite{B2009.1,MS2013.1,MS2014.2}
\begin{equation}
\label{SP off off}
\fl\hspace{20mm}
\langle\psi_{\vec{w}}(\gamma)|\psi_{\vec{y}}(\gamma)\rangle=\Big(\prod_{j=1}^{N}\frac{(1-y_{j})\,w_{j}^{N}}{y_{j}\,(1-w_{j})^{L}}\Big)\,\det\Bigg(\frac{\frac{(1-w_{k})^{L}}{w_{k}^{N-1}}-\frac{(1-y_{j})^{L}}{y_{j}^{N-1}}}{y_{j}-w_{k}}\Bigg)_{j,k\in\[1,N\]}\;.
\end{equation}
Taking the $y_{j}$ in (\ref{SP off off}) as solutions of the Bethe equations (\ref{Bethe equations}) with fugacity $\gamma$ while keeping the $w_{k}$ generic, the off-shell / on-shell scalar product reduces to the Slavnov determinant \cite{S1989.1}
\begin{eqnarray}
\label{SP off on}
&&\fl\hspace{10mm}
\langle\psi_{\vec{w}}(\gamma)|\psi_{\vec{y}}(\gamma)\rangle=(-1)^{N}\Big(\prod_{j=1}^{N}\frac{(1-y_{j})^{L+1}}{y_{j}^{N}(1-w_{j})^{L}}\Big)\Bigg(\prod_{j=1}^{N}\prod_{k=1}^{N}(y_{j}-w_{k})\Bigg)\\
&&\hspace{9mm}
\times\det\Bigg(\partial_{y_{i}}\Big(\prod_{k=1}^{N}\frac{1}{1-y_{k}/w_{j}}+\rme^{L\gamma}(1-w_{j})^{L}\prod_{k=1}^{N}\frac{1}{1-w_{j}/y_{k}}\Big)\Bigg)_{i,j\in\[1,N\]}\;,\nonumber
\end{eqnarray}
where the derivative in the determinant has to be taken before setting the $y_{j}$ to a solution of the Bethe equations. Finally, taking the singular limit $w_{j}\to y_{j}$ in (\ref{SP off on}), the on-shell norm of the Bethe vector is given by the Gaudin determinant \cite{GMCW1981.1,K1982.1,B2009.1,MSS2012.2}
\begin{equation}
\label{SP on on}
\langle\psi_{\vec{y}}(\gamma)|\psi_{\vec{y}}(\gamma)\rangle=\frac{L}{N}\Bigg(\sum_{j=1}^{N}\frac{y_{j}}{N+(L-N)y_{j}}\Bigg)\Bigg(\prod_{j=1}^{N}\frac{N+(L-N)y_{j}}{y_{j}}\Bigg)\;,
\end{equation}
in terms of which one has the resolution of the identity
\begin{equation}
\label{1[psi]}
\mathbf{1}=\sum_{r=1}^{|\Omega_{L,N}|}\frac{|\psi_{r}(\gamma)\rangle\langle\psi_{r}(\gamma)|}{\langle\psi_{r}(\gamma)|\psi_{r}(\gamma)\rangle},
\end{equation}
with the basis $\psi_{r}(\gamma)$, $r=1,\ldots,|\Omega_{L,N}|$ corresponding to all admissible solutions of the Bethe equations.

Additionally, in order to expand current fluctuations of TASEP over Bethe eigenstates in section~\ref{section Height fluctuations}, one needs to consider modified Bethe vectors
\begin{eqnarray}
\label{psi0 r}
&& \langle x_{1},\ldots,x_{N}|\psi_{\vec{y}}^{0}\rangle=\det(y_{j}^{-k}(1-y_{j})^{x_{k}})_{j,k\in\[1,N\]}\\
\label{psi0 l}
&& \langle\psi_{\vec{y}}^{0}|x_{1},\ldots,x_{N}\rangle=\det(y_{j}^{k}(1-y_{j})^{-x_{k}})_{j,k\in\[1,N\]}\;,
\end{eqnarray}
which are eigenstates of a deformed Markov operator $M_{0}(L\gamma)$ counting the current of particles between sites $L$ and $1$ if the $y_{j}$ are solution of the Bethe equations (\ref{Bethe equations}), see section~\ref{section Height fluctuations}. One has the identity \cite{B2009.1,P2016.1}
\begin{equation}
\sum_{\mathcal{C}\in\Omega_{L,N}}\langle\mathcal{C}|\psi_{\vec{y}}^{0}\rangle
=(1-\rme^{-L\gamma})\Big(\prod_{j=1}^{N}\frac{1-y_{j}}{y_{j}^{N+1}}\Big)\Big(\prod_{j=1}^{N}\prod_{k=j+1}^{N}(y_{j}-y_{k})\Big)
\end{equation}
for Bethe roots $y_{j}$ solutions of the Bethe equations with fugacity $\gamma$. Furthermore, the on-shell scalar product for two sets of Bethe roots $w_{j}$ and $y_{j}$ solution of the Bethe equation with respective fugacity $\gamma_{w}$ and $\gamma_{y}$ is equal to
\begin{eqnarray}
\label{SP psi0 psi0}
&&\fl\hspace{15mm}
\langle\psi^{0}_{\vec{w}}|\psi^{0}_{\vec{y}}\rangle=(-1)^{\frac{N(N-1)}{2}}
\Big(\prod_{j=1}^{N}\frac{(1-y_{j})w_{j}}{y_{j}}\Big)
\Big(1-\frac{\rme^{L\gamma_{w}}}{\rme^{L\gamma_{y}}}\Big)
\Big(1-\frac{\rme^{L\gamma_{w}}\prod_{j=1}^{N}w_{j}}{\rme^{L\gamma_{y}}\prod_{j=1}^{N}y_{j}}\Big)^{N-1}\nonumber\\
&&\hspace{25mm}
\times\frac{(\prod_{j=1}^{N}\prod_{k=j+1}^{N}(y_{j}-y_{k}))(\prod_{j=1}^{N}\prod_{k=j+1}^{N}(w_{j}-w_{k}))}{\prod_{j=1}^{N}\prod_{k=1}^{N}(y_{j}-w_{k})}\;.
\end{eqnarray}
A derivation is provided in \ref{appendix SP psi0 psi0}, see also \cite{BL2019.1}, proposition~5.2 for an alternative proof which does not use the Slavnov determinant formula.
\end{subsection}

\end{section}

%%%%%%%%%%%%%%%%%%%
%%               %%
%%  Section III  %%
%%               %%
%%%%%%%%%%%%%%%%%%%
\begin{section}{Riemann surfaces}
\label{section Riemann surfaces}
In this section, we introduce meromorphic Bethe root functions $y_{j}(C)$, $j\in\[1,N\]$, whose domain can be extended by analytic continuation to a Riemann surface $\R_{1}$ isomorphic to the Riemann sphere $\Ch$. Then, we consider symmetric functions of $N$ Bethe roots, and the corresponding Riemann surface $\RN$ used for eigenvalues and eigenvectors of TASEP in section~\ref{section Height fluctuations}. We refer to \cite{B2013.3,E2018.1} for an introduction to compact Riemann surfaces.

\begin{subsection}{Polynomial equation and generalized Cassini ovals}
\label{section Cassini}
Let $(y_{1},\ldots,y_{N})$ be a solution of the Bethe equations (\ref{Bethe equations}) with fugacity $\gamma$. Introducing the parameter \footnote{The extra factor $\rho^{N}(1-\rho)^{L-N}$ is included in preparation for the large $L$ limit.}
\begin{equation}
\label{C[y]}
C=\frac{\rme^{L\gamma}}{\rho^{N}(1-\rho)^{L-N}}\prod_{k=1}^{N}y_{k}
\end{equation}
with $\rho=N/L$ the average density of particles, the Bethe equations rewrite as the polynomial equation $P(y_{j},C)=0$, $j\in\[1,N\]$ with
\begin{equation}
\label{P}
P(y,C)=\rho^{N}(1-\rho)^{L-N}C(1-y)^{L}+(-1)^{N}y^{N}\;.
\end{equation}
For a given value of $|C|$, all $L$ solutions $y$ of $P(y,C)=0$ belong to the generalized Cassini oval \cite{GM2005.1} $\rho^{N}(1-\rho)^{L-N}|C|\,|1-y|^{L}=|y|^{N}$ plotted in figure~\ref{fig Cassini}.

\begin{figure}
	\begin{tabular}{lll}
		\begin{tabular}{l}
			\includegraphics[width=42mm]{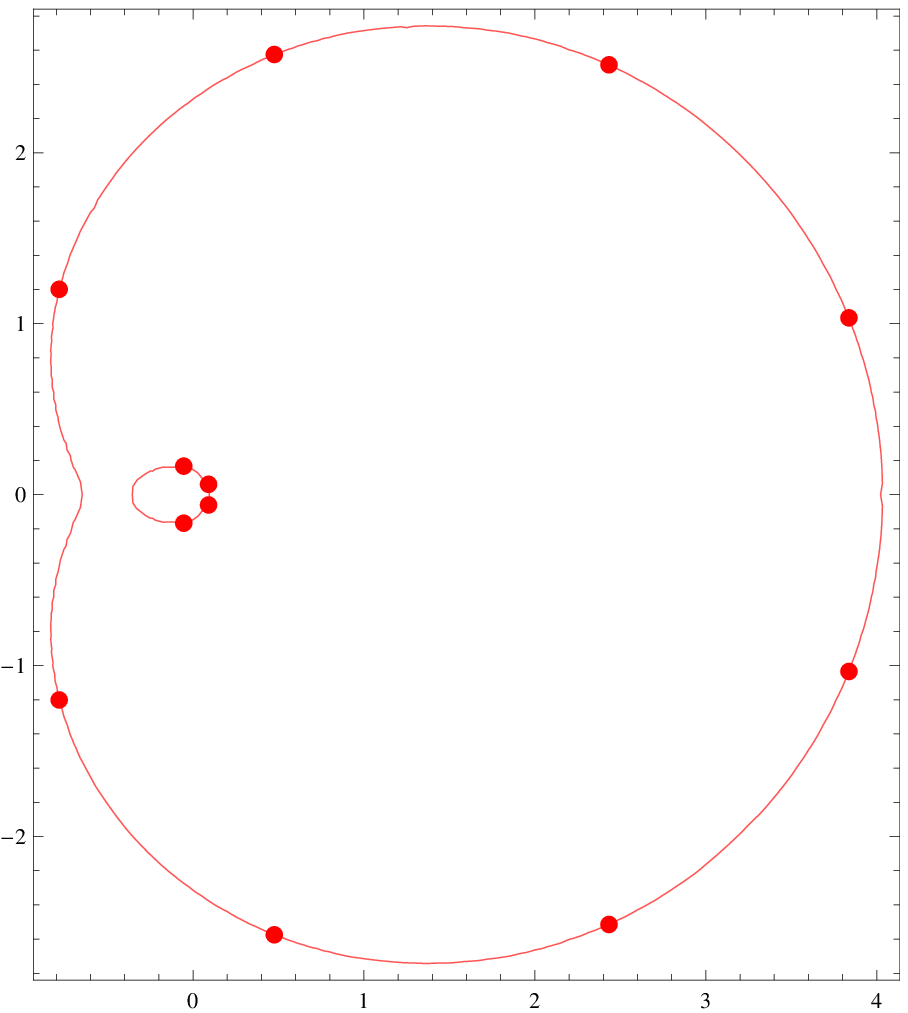}
			\begin{picture}(0,0)
				\put(-29,35){\small$|C|<1$}
			\end{picture}
		\end{tabular}
		&
		\begin{tabular}{l}
			\includegraphics[width=42mm]{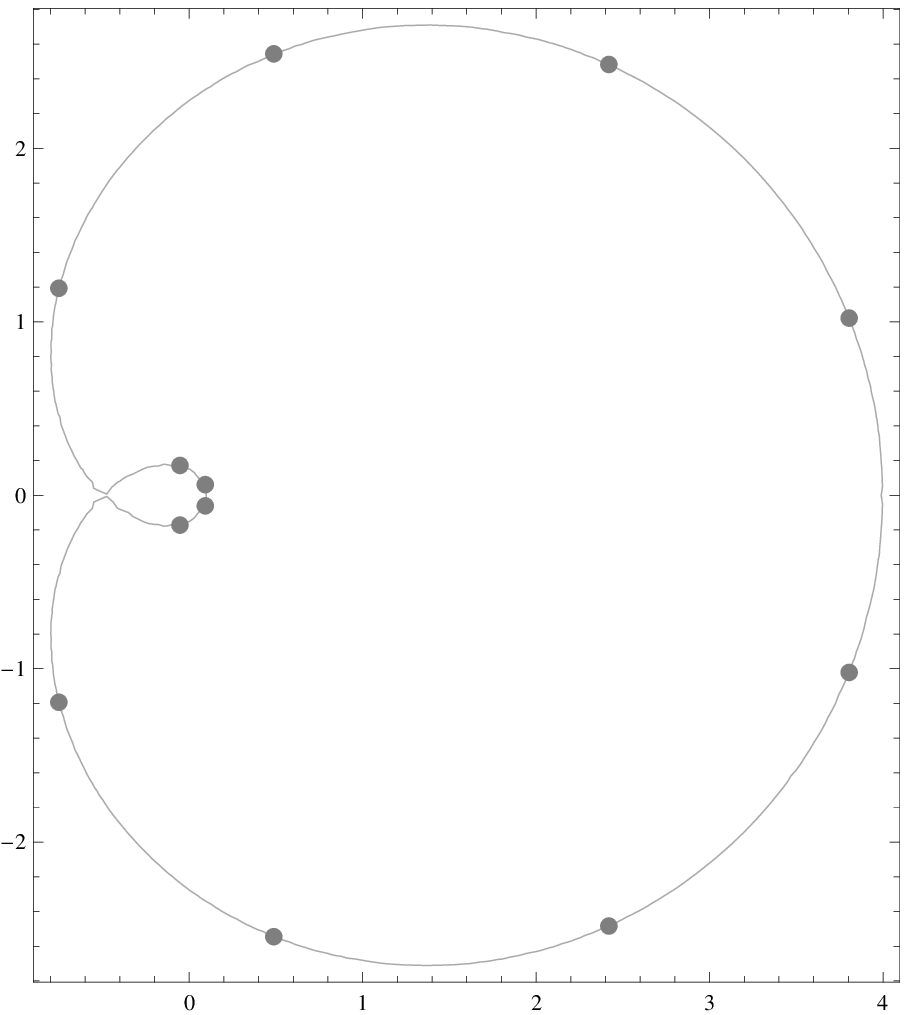}
			\begin{picture}(0,0)
				\put(-29,35){\small$|C|=1$}
			\end{picture}
		\end{tabular}
		&
		\begin{tabular}{l}
			\includegraphics[width=42mm]{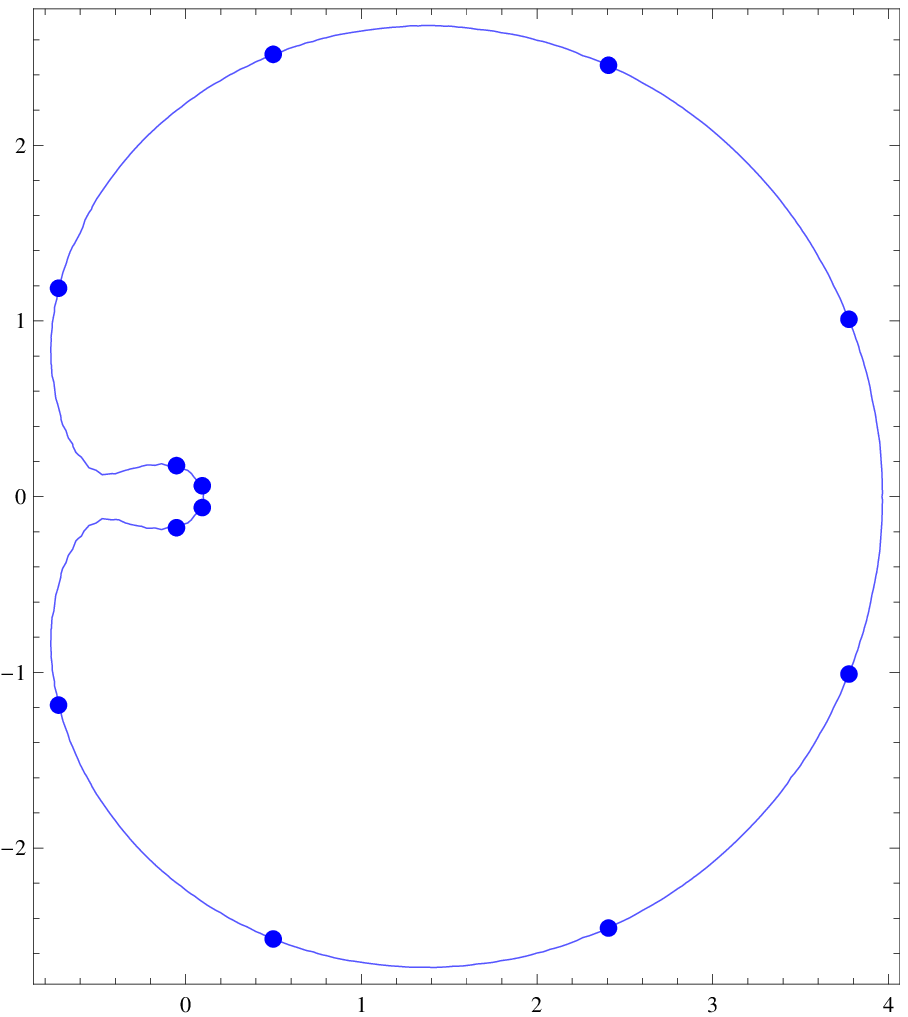}
			\begin{picture}(0,0)
				\put(-29,35){\small$|C|>1$}
			\end{picture}
		\end{tabular}
	\end{tabular}
	\caption{Generalized Cassini oval for $\rho=1/3$ and $|C|=0.9$ (left), $|C|=1$ (middle) and $|C|=1.1$ (right). The dots represent the corresponding solutions $y\in\mathbb{C}$ of $P(y,C)$ for $L=12$, $N=4$ and $C>0$.}
	\label{fig Cassini}
\end{figure}

When $C\to0$, $N$ solutions $y_{j}$ of $P(y_{j},C)=0$ converge to $0$ as $C^{1/N}$, the remaining $L-N$ solutions diverge as $C^{-1/(L-N)}$, and the Cassini oval is composed of two disjoint circles. When $C\to\infty$ on the other hand, all $L$ solutions converge to $1$ as $1-y_{j}\sim C^{-1/L}$, and the Cassini oval is a single circle around $1$. The transition from the small $|C|$ (two disjoint closed curves) to the large $|C|$ (a single closed curve) behaviour necessarily occurs at a value of $|C|$ for which the two disjoint curves intersect. Thus, there must exist $C^{*}$ such that the equation $P(y,C^{*})=0$ has a double root $y^{*}$, corresponding to the existence of a branch point at $C=C^{*}$ for some branch of the multivalued function $y(C)$ solution of $P(y(C),C)=0$. The pair $(C^{*},y^{*})$ must then be a solution of the system $P(y^{*},C^{*})=0$, $\partial_{y}P(y^{*},C^{*})=0$, whose unique solution with $C^{*}\notin\{0,\infty\}$ is $C^{*}=-1$, $y^{*}=-\frac{\rho}{1-\rho}$. One can conclude that the generalized Cassini oval is made of two connected components when $|C|<1$, which merge for $|C|=1$ at $y=-\frac{\rho}{1-\rho}$, so that the curve has a single connected component when $|C|>1$, see figure~\ref{fig Cassini}.
\end{subsection}

\begin{subsection}{Bethe root functions \texorpdfstring{$y_{j}(C)$}{yj(C)} on \texorpdfstring{$\D$}{D}}
\label{section yj(C)}
\begin{figure}
	\begin{center}
		\begin{tabular}{c}
			\includegraphics[width=150mm]{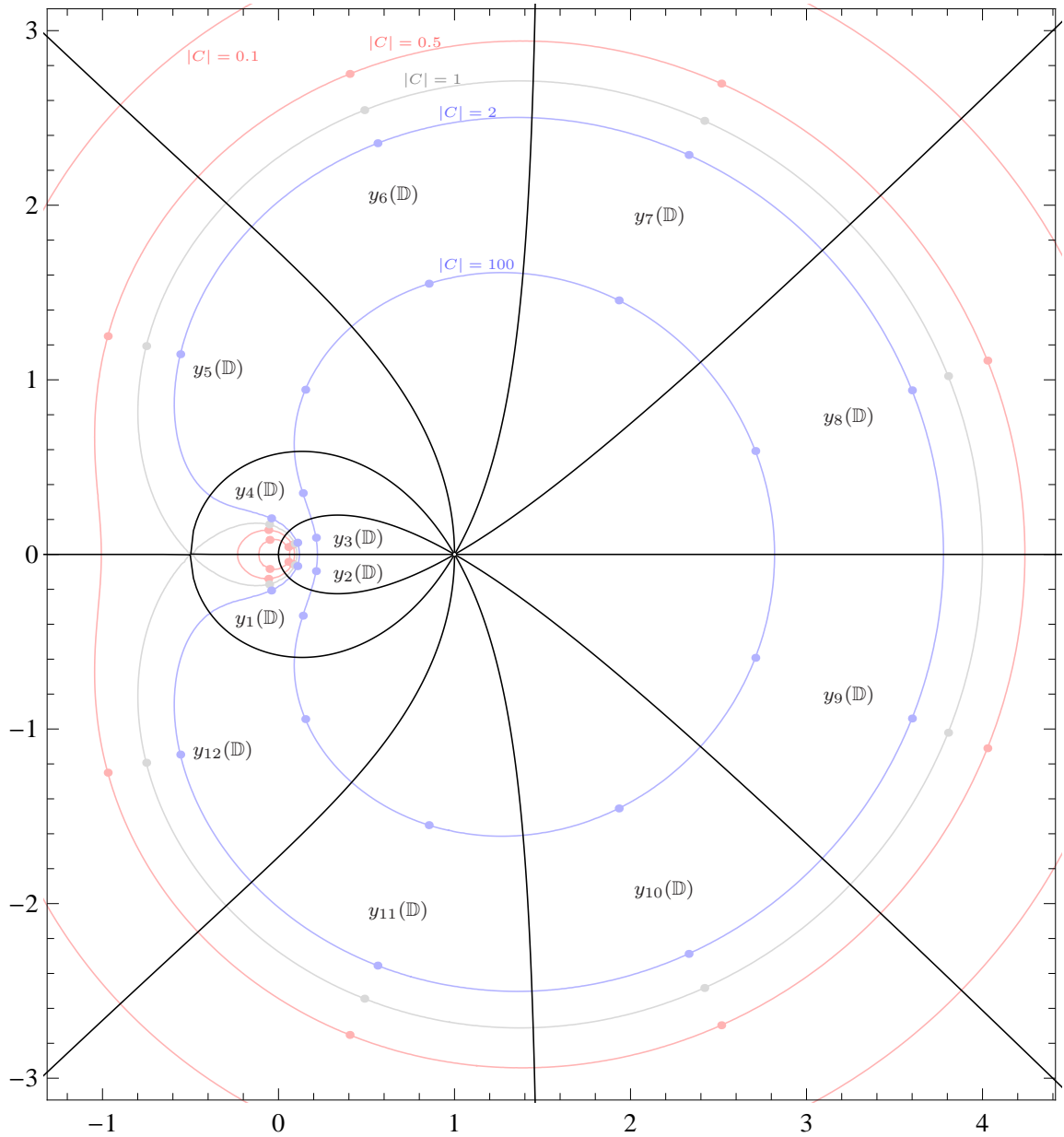}
			\begin{picture}(0,0)
				\put(-119,74){\scriptsize$y_{1}(\D)$}
				\put(-105,80.5){\scriptsize$y_{2}(\D)$}
				\put(-105,85.5){\scriptsize$y_{3}(\D)$}
				\put(-119,92.5){\scriptsize$y_{4}(\D)$}
				\put(-125,110){\scriptsize$y_{5}(\D)$}
				\put(-100,135){\scriptsize$y_{6}(\D)$}
				\put(-62,132){\scriptsize$y_{7}(\D)$}
				\put(-35,103){\scriptsize$y_{8}(\D)$}
				\put(-35,63){\scriptsize$y_{9}(\D)$}
				\put(-62,35){\scriptsize$y_{10}(\D)$}
				\put(-100,32){\scriptsize$y_{11}(\D)$}
				\put(-125,55){\scriptsize$y_{12}(\D)$}
				\put(-126,155){\tiny\color[rgb]{1,0.4,0.4}$|C|=0.1$}
				\put(-100,157){\tiny\color[rgb]{1,0.4,0.4}$|C|=0.5$}
				\put(-95,151.7){\tiny\color[rgb]{0.5,0.5,0.5}$|C|=1$}
				\put(-90,147){\tiny\color[rgb]{0.4,0.4,1}$|C|=2$}
				\put(-90,125){\tiny\color[rgb]{0.4,0.4,1}$|C|=100$}
			\end{picture}
		\end{tabular}
	\end{center}
	\caption{Domains $y_{j}(\D)$ for $L=12$, $N=4$. The domains are delimited by black curves. Intersections of black curves correspond to ramification points of the covering map $\pi_{1}$. The lighter curves represent the generalized Cassini ovals for $\rho=1/3$ with $|C|\in\{0.1,0.5,1,2,100\}$, and the dots are the corresponding solutions of $P(y,C)=0$ with $C>0$.}
	\label{fig yj}
\end{figure}

When $C\notin\{0,-1,\infty\}$, the equation $P(y,C)=0$ has $L$ distinct solutions $y$. We label these solutions as functions $y_{j}(C)$, $j\in\[1,L\]$ analytic in $\D=\mathbb{C}\setminus\mathbb{R}^{-}$ as in figure~\ref{fig yj}: $y_{1}(C),\ldots,y_{N}(C)$ are bounded for $C\in\D$ and ordered as $-\pi<\arg y_{1}(C)<\ldots<\arg y_{N}(C)<\pi$ while $y_{N+1}(C),\ldots,y_{L}(C)$ are unbounded for $C\in\D$ and ordered as $\pi>\arg y_{N+1}(C)>\ldots>\arg y_{L}(C)>-\pi$. The small and large $C$ behaviours of the $y_{j}(C)$ are then given by
\begin{equation}
\label{yj C0}
\fl\hspace{10mm}
y_{j}(C)\underset{C\to0}{\simeq}
\Bigg\{
\begin{array}{lll}
\rme^{\frac{2\rmi\pi}{N}\big(j-\frac{N+1}{2}\big)}\,\rho\,(1-\rho)^{\frac{1-\rho}{\rho}}\,C^{1/N} && 1\leq j\leq N\\[2mm]
\rme^{-\frac{2\rmi\pi}{L-N}\big(j-\frac{L+N+1}{2}\big)}\,\rho^{-\frac{\rho}{1-\rho}}\,(1-\rho)^{-1}\,C^{-1/(L-N)} && N+1\leq j\leq L
\end{array}
\end{equation}
and
\begin{equation}
\label{yj Cinf}
1-y_{j}(C)\underset{C\to\infty}{\simeq}\rme^{-\frac{2\rmi\pi}{L}\big(j-\frac{N+1}{2}\big)}\,\rho^{-\rho}\,(1-\rho)^{-(1-\rho)}\,C^{-1/L}\;,
\end{equation}
where fractional powers are defined with the branch cut $\mathbb{R}^{-}$.

\begin{figure}
	\begin{center}
		\begin{picture}(150,50)
			\put(0,25){\color{red}\linethickness{0.5mm}\line(1,0){75}}
			\put(37.5,25){\color{red}\circle*{2}}
			\put(75,25){\color{red}\circle*{2}}
			\put(35,20){$-1$}
			\put(74,20){$0$}
			\put(20,50){\thicklines\vector(0,-1){20}}
			\put(55,50){\thicklines\vector(0,-1){20}}
			\put(20,0){\thicklines\vector(0,1){20}}
			\put(55,0){\thicklines\vector(0,1){20}}
			\put(11,40){$\mathcal{A}_{\text{out}}$}
			\put(56,40){$\mathcal{A}_{\text{in}}$}
			\put(11,9){$\mathcal{A}_{\text{out}}^{-1}$}
			\put(56,9){$\mathcal{A}_{\text{in}}^{-1}$}
		\end{picture}
	\end{center}
	\caption{Analytic continuation operators across the cuts $(-\infty,-1)$ and $(-1,0)$.}
	\label{fig a.c.}
\end{figure}
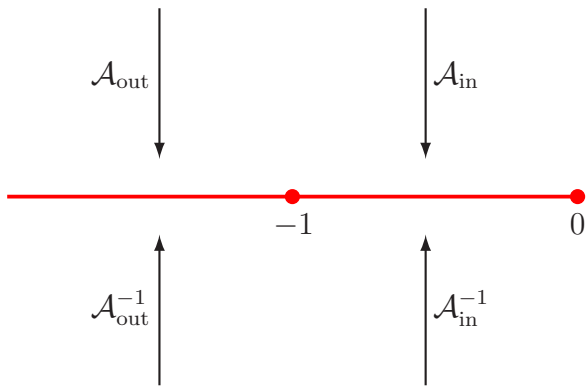

Increasing $\arg C$ while keeping $|C|$ fixed, the points $y_{j}(C)$ move on the generalized Cassini ovals in the counter-clockwise direction when $|C|<1$ and $j\in\[1,N\]$, and in the clockwise direction when either $|C|>1$ or $|C|<1$ and $j\in\[N+1,L\]$. Starting with a function $y_{j}$, the analytic continuation across either branch cut $(-\infty,-1)$ or $(-1,0)$ leads to a function $y_{k}$ also analytic in $\D$. Depending on which side the branch cut is crossed, four distinct values of $k$ are possible. We write $y_{k}=\mathcal{A}_{\text{in}}y_{j}$ or $y_{k}=\mathcal{A}_{\text{out}}y_{j}$ respectively if the branch cut $(-1,0)$ or $(-\infty,-1)$ is crossed from above, and $y_{k}=\mathcal{A}_{\text{in}}^{-1}y_{j}$ and $y_{k}=\mathcal{A}_{\text{out}}^{-1}y_{j}$ if the cuts are crossed from below, see figure~\ref{fig a.c.}. This defines bijections $A_{\text{in}}$, $A_{\text{out}}$ on $\[1,L\]$ such that $\mathcal{A}_{\text{in}}y_{j}=y_{A_{\text{in}}j}$ and $\mathcal{A}_{\text{out}}y_{j}=y_{A_{\text{out}}j}$. One has
\begin{equation}
\label{Aout}
\begin{array}{lll}
A_{\text{out}}j=j+1 && 1\leq j<L\\
A_{\text{out}}L=1 && j=L
\end{array}
\end{equation}
and
\begin{equation}
\label{Ain}
\begin{array}{lll}
A_{\text{in}}j=j+1 && 1\leq j<N\\
A_{\text{in}}N=1 && j=N\\
A_{\text{in}}j=j+1 && N+1\leq j<L\\
A_{\text{in}}L=N+1 && j=L
\end{array}\;.
\end{equation}
The operators $A_{\text{in}}$ and $A_{\text{out}}$ generate a subgroup $G$ of the permutation group of $\[1,L\]$, with cardinal $|G|=\gcd(L,N)(\frac{L}{\gcd(L,N)}!)^{\gcd(L,N)}$, and equal to the full symmetric group if and only if $L$ and $N$ are co-prime.

For later reference, we note that the derivative of the function $y_{j}$ is given by
\begin{equation}
\label{yj'}
y_{j}'(C)=\frac{1}{C}\,\frac{y_{j}(C)\,(1-y_{j}(C))}{N+(L-N)\,y_{j}(C)}\;.
\end{equation}
\end{subsection}

\begin{subsection}{Riemann sphere \texorpdfstring{$\R_{1}\sim\Ch$}{R1\~Chat}}
The compact Riemann surface $\R_{1}$ obtained by gluing together along the cuts $(-\infty,-1)$, $(-1,0)$ the domains of definition of the functions $y_{j}$ according to analytic continuations is composed of $L$ sheets. The points of $\R_{1}$ may be labelled as $[C,j]$, $j\in\[1,L\]$, $C\in\Ch$, where $\Ch=\C\cup\{\infty\}$ is the Riemann sphere, and the functions $y_{j}$ may then be extended to a meromorphic function $y:\R_{1}\to\Ch$ by
\begin{equation}
\label{y[yj]}
y([C,j])=y_{j}(C)\;,
\end{equation}
whose analytic properties are discussed toward the end of this section. In order to discuss some features of the Riemann surface $\R_{1}$, we define the covering map $\pi_{1}:\R_{1}\to\Ch$ by $\pi_{1}([C,j])=C$, which has degree $L$ (number of antecedents of a generic point from the target Riemann surface $\Ch$).

We recall that ramification points of a covering map $\pi:\mathcal{M}\to\mathcal{N}$ between Riemann surfaces $\mathcal{M}$ and $\mathcal{N}$ are the $p\in\mathcal{M}$ such that a small closed circle around $\pi(p)\in\mathcal{N}$ does not pull back under $\pi$ to a closed path around $p$, and the image $\pi(p)$ of a ramification point by the covering map is called a branch point. The ramification index $e_{p}\geq2$ of a ramification point is the smallest positive winding number around $\pi(p)$ of a closed path in a neighbouring of $\pi(p)$ that pulls back to a closed curve around $p$.

\begin{table}
	\begin{center}
		\begin{tabular}{lcc}
			\hspace{12mm}Ramification point $p\in\R_{1}$ & \begin{tabular}{c}Ramification\\index $e_{p}$\end{tabular} & \begin{tabular}{c}Branch point\\$\pi_{1}(p)\in\Ch$\end{tabular}\\[6mm]
			\hspace{9.5mm}$y^{-1}(0)=[0,1]=\ldots=[0,N]$ & $N$ & $0$\\[3mm]
			\hspace{7.5mm}$y^{-1}(\infty)=[0,N+1]=\ldots=[0,L]$ & $L-N$ & $0$\\[3mm]
			\hspace{9.5mm}$y^{-1}(1)=[\infty,1]=\ldots=[\infty,L]$ & $L$ & $\infty$\\[3mm]
			\begin{tabular}{l}$y^{-1}(-\tfrac{\rho}{1-\rho})=[-1-\rmi\epsilon,1]=[-1+\rmi\epsilon,N]$\\\hspace{20mm}$=[-1-\rmi\epsilon,N+1]=[-1+\rmi\epsilon,L]$\end{tabular} & $2$ & $-1$
		\end{tabular}
	\end{center}
	\caption{List of all four ramification points of the ramified covering $\pi_{1}:\R_{1}\to\Ch$.}
	\label{table ramif pi1}
\end{table}

The branch points of $\pi_{1}$ are all the possible branch points $0$, $-1$, $\infty$ of the functions $y_{j}$ from which $\R_{1}$ was built. By construction of $\R_{1}$, small closed paths around $[0,j]$ are generated by repeated action of $A_{\text{in}}$ on $j$, see figure~\ref{fig a.c.}, and the two ramification points of $\pi_{1}$ corresponding to the branch point $0\in\Ch$ are $[0,1]=\ldots=[0,N]$ with ramification index $N$ and $[0,N+1]=\ldots=[0,L]$ with ramification index $L-N$. Similarly, small closed paths around $[\infty,j]$ are generated by repeated action of $A_{\text{out}}$ on $j$, and the branch point $\infty\in\Ch$ corresponds to the single ramification point $[\infty,1]=\ldots=[\infty,L]\in\R_{1}$ with ramification index $L$. Finally since $(A_{\text{in}}^{-1}A_{\text{out}})^{2}$ is the identity permutation, small paths around $-1$ with winding number $2$ always lift by $\pi_{1}^{-1}$ to closed loops on $\R_{1}$, and the elements of $\pi_{1}^{-1}(-1)$ are either regular points or ramification points with ramification index $2$. Since $A_{\text{in}}^{-1}A_{\text{out}}$ is the transposition between $N$ and $L$ and $A_{\text{out}}A_{\text{in}}^{-1}$ the transposition between $1$ and $N+1$, we find that all the antecedents of $-1$ are regular points with respect to $\pi_{1}$ except for $[-1-\rmi\epsilon,1]=[-1+\rmi\epsilon,N]=[-1-\rmi\epsilon,N+1]=[-1+\rmi\epsilon,L]$, $0<\epsilon\to0$, which is a ramification point with ramification index $2$.

All the points $[C,j]$, $C\in\Ch$, $j\in\[1,L\]$ of $\R_{1}$ are distinct, except for the identifications discussed above at ramification points of $\pi_{1}$, visible on figure~\ref{fig yj} as intersections of black curves, and which are summarized in table~\ref{table ramif pi1}.

The genus of $\R_{1}$ is equal to $0$, as is easily seen on figure~\ref{fig yj} after compactification by adding the point at infinity. This can also be obtained from the Riemann-Hurwitz formula, which reads for a covering map $\pi:\mathcal{M}\to\mathcal{N}$ of degree $d$ between connected, compact Riemann surfaces $\mathcal{M}$ and $\mathcal{N}$ of respective genus $g_{\mathcal{M}}$ and $g_{\mathcal{N}}$ as
\begin{equation}
\label{Riemann Hurwitz}
g_{\mathcal{M}}=d(g_{\mathcal{N}}-1)+1+\frac{1}{2}\sum_{p\in\mathcal{M}}(e_{p}-1)\;.
\end{equation}
For the covering map $\pi_{1}$, calling $g_{1}$ the genus of $\R_{1}$, one finds indeed $g_{1}=-L+1+((N-1)+(L-N-1)+(L-1)+(2-1))/2=0$, see table~\ref{table ramif pi1}. Since the Riemann sphere is the only Riemann surface of genus $0$ up to isomorphism, one has
\begin{equation}
\R_{1}\sim\Ch\;.
\end{equation}

From the small $C$ behaviour of the solutions of $P(y,C)=0$ discussed in section~\ref{section Cassini}, see also figure~\ref{fig yj}, the function $y:\R_{1}\to\Ch$ defined in (\ref{y[yj]}) has the single pole $[0,N+1]=\ldots=[0,L]$ on $\R_{1}$, and $y$ is thus bijective since, by a general property of non-constant meromorphic functions on compact Riemann surfaces, the point $\infty\in\Ch$ must have the same number of antecedents as any other point in $\Ch$. The four ramification points of $\pi_{1}$ discussed above can be identified as $y^{-1}(0)$, $y^{-1}(\infty)$, $y^{-1}(1)$ and $y^{-1}(-\frac{\rho}{1-\rho})$, see table~\ref{table ramif pi1}. Equivalently, $0$ is a branch point of the $y_{j}$, of order $N$ for $1\leq j\leq N$ and of order $L-N$ for $N+1\leq j\leq L$, $\infty$ is a branch point of order $L$ for all $y_{j}$, and $-1$ is a branch point of order $2$ (square root branch point) for $y_{1}$, $y_{N}$, $y_{N+1}$ and $y_{L}$. Since $0$ and $\infty$ are extremities of the cut $\mathbb{R}^{-}$ of $\D$ on which the functions $y_{j}$ are defined, the branch point is reached by approaching $0$ or $\infty$ from any direction. This is not the case for the point $-1$, which can be approached either from above ($C=-1+\rmi\epsilon$, $0<\epsilon\to0$) or from below ($C=-1-\rmi\epsilon$, $0<\epsilon\to0$) the cut. The point $-1$ is a branch point for $y_{1}$ and $y_{N+1}$ only (respectively for $y_{N}$ and $y_{L}$ only) when it is approached from below (resp. from above).

The Riemann surface $\R_{1}$ can alternatively be constructed directly from the algebraic curve $P(y,C)=0$ after a desingularization procedure at the conical singularities $C=0$ and $C=\infty$, where all the sheets are connected in the algebraic curve but not in the Riemann surface.
\end{subsection}

\begin{subsection}{Symmetric functions of $N$ Bethe roots and Riemann surface \texorpdfstring{$\RN$}{RN}}
Let us consider an arbitrary symmetric meromorphic function $s$ in $N$ variables, and a subset $J=\{j_{1},\ldots,j_{N}\}$ of $\[1,L\]$ with $|J|=N$ elements. The function of one variable $s_{J}:C\mapsto s(y_{j_{1}}(C),\ldots,y_{j_{N}}(C))$ with $y_{j}(C)$ defined as in section~\ref{section yj(C)} is meromorphic in $\D=\C\setminus\mathbb{R}^{-}$, and we are interested in the compact Riemann surface $\RN$ to which $s_{J}$ can be extended by analytic continuations.

Crossing the cut $(-1,0)$ (respectively $(-\infty,-1)$) from above, $s_{J}$ is continued analytically to $s_{A_{\text{in}}J}$ (resp. $s_{A_{\text{out}}J}$), where the operators $A_{\text{in}}$, $A_{\text{out}}$ are extended to sets of $N$ indices by $A_{\text{in}}\{j_{1},\ldots,j_{N}\}=\{A_{\text{in}}\,j_{1},\ldots,A_{\text{in}}\,j_{N}\}$ and $A_{\text{out}}\{j_{1},\ldots,j_{N}\}=\{A_{\text{out}}\,j_{1},\ldots,A_{\text{out}}\,j_{N}\}$. The Riemann surface $\RN$ is constructed by gluing together according to analytic continuations all ${{L}\choose{N}}$ sheets on which the functions $s_{J}$ live. The points of $\RN$ are written as $[C,J]$, $C\in\Ch$ with $J\subset\[1,L\]$, $|J|=N$ indexing the sheets \footnote{In the following, the sheets of $\RN$ and the sets $J$ are identified by abuse of language.}, and the ramified covering $\pi_{N}:\RN\to\Ch$ defined by $\pi_{N}[C,J]=C$ is of degree ${{L}\choose{N}}$.

\begin{table}
\begin{center}
	\begin{tabular}{r|ccccccccccccccc}
	$L\;\backslash\;N$ & 1 & 2 & 3 & 4 & 5 & 6 & 7 & 8 & 9 & 10 & 11 & 12 & 13 & 14 & 15\\\hline
	2 & 1 & $\cdot$ & $\cdot$ & $\cdot$ & $\cdot$ & $\cdot$ & $\cdot$ & $\cdot$ & $\cdot$ & $\cdot$ & $\cdot$ & $\cdot$ & $\cdot$ & $\cdot$ & $\cdot$\\
	3 & 1 & 1 & $\cdot$ & $\cdot$ & $\cdot$ & $\cdot$ & $\cdot$ & $\cdot$ & $\cdot$ & $\cdot$ & $\cdot$ & $\cdot$ & $\cdot$ & $\cdot$ & $\cdot$\\
	4 & 1 & 2 & 1 & $\cdot$ & $\cdot$ & $\cdot$ & $\cdot$ & $\cdot$ & $\cdot$ & $\cdot$ & $\cdot$ & $\cdot$ & $\cdot$ & $\cdot$ & $\cdot$\\
	5 & 1 & 1 & 1 & 1 & $\cdot$ & $\cdot$ & $\cdot$ & $\cdot$ & $\cdot$ & $\cdot$ & $\cdot$ & $\cdot$ & $\cdot$ & $\cdot$ & $\cdot$\\
	6 & 1 & 2 & 3 & 2 & 1 & $\cdot$ & $\cdot$ & $\cdot$ & $\cdot$ & $\cdot$ & $\cdot$ & $\cdot$ & $\cdot$ & $\cdot$ & $\cdot$\\
	7 & 1 & 1 & 1 & 1 & 1 & 1 & $\cdot$ & $\cdot$ & $\cdot$ & $\cdot$ & $\cdot$ & $\cdot$ & $\cdot$ & $\cdot$ & $\cdot$\\
	8 & 1 & 2 & 1 & 6 & 1 & 2 & 1 & $\cdot$ & $\cdot$ & $\cdot$ & $\cdot$ & $\cdot$ & $\cdot$ & $\cdot$ & $\cdot$\\
	9 & 1 & 1 & 4 & 1 & 1 & 4 & 1 & 1 & $\cdot$ & $\cdot$ & $\cdot$ & $\cdot$ & $\cdot$ & $\cdot$ & $\cdot$\\
	10 & 1 & 2 & 1 & 3 & 11 & 3 & 1 & 2 & 1 & $\cdot$ & $\cdot$ & $\cdot$ & $\cdot$ & $\cdot$ & $\cdot$\\
	11 & 1 & 1 & 1 & 1 & 1 & 1 & 1 & 1 & 1 & 1 & $\cdot$ & $\cdot$ & $\cdot$ & $\cdot$ & $\cdot$\\
	12 & 1 & 2 & 4 & 9 & 1 & 26 & 1 & 9 & 4 & 2 & 1 & $\cdot$ & $\cdot$ & $\cdot$ & $\cdot$\\
	13 & 1 & 1 & 1 & 1 & 1 & 1 & 1 & 1 & 1 & 1 & 1 & 1 & $\cdot$ & $\cdot$ & $\cdot$\\
	14 & 1 & 2 & 1 & 3 & 1 & 4 & 57 & 4 & 1 & 3 & 1 & 2 & 1 & $\cdot$ & $\cdot$\\
	15 & 1 & 1 & 4 & 1 & 21 & 9 & 1 & 1 & 9 & 21 & 1 & 4 & 1 & 1 & $\cdot$\\
	16 & 1 & 2 & 1 & 10 & 1 & 4 & 1 & 142 & 1 & 4 & 1 & 10 & 1 & 2 & 1
	\end{tabular}
\end{center}
\caption{Number of connected components of the Riemann surface $\RN$ for small values of $L$ and $N$.}
\label{table nb cc RN}
\end{table}

Orbits under the action on subsets of $N$ elements of $\[1,L\]$ of the group $G$ generated by $A_{\text{in}}$, $A_{\text{out}}$ correspond to connected components of $\RN$. Defining $M=\gcd(L,N)$, two sheets $J$ and $K$ belong to the same connected component if and only if there exists $m\in\Z/M\Z$ such that $K=J+m$ modulo $M$. The connected component $\RNp$ of $\RN$ containing the \emph{principal sheet} $\[1,N\]$ is called in the following the \emph{principal connected component} of $\RN$. The number of connected components of $\RN$, given in table~\ref{table nb cc RN} for small values of $L,N$, is invariant under $(L,N)\to(L,L-N)$, and is equal to $1$ if and only if $L$ and $N$ are co-prime, in which case $\RN=\RNp$.

\begin{table}
\begin{center}
	\begin{tabular}{r|ccccccccccccccc}
	$L\;\backslash\;N$ & 1 & 2 & 3 & 4 & 5 & 6 & 7 & 8 & 9 & 10 & 11 & 12 & 13 & 14 & 15\\\hline
	2 & 0 & $\cdot$ & $\cdot$ & $\cdot$ & $\cdot$ & $\cdot$ & $\cdot$ & $\cdot$ & $\cdot$ & $\cdot$ & $\cdot$ & $\cdot$ & $\cdot$ & $\cdot$ & $\cdot$\\
	3 & 0 & 0 & $\cdot$ & $\cdot$ & $\cdot$ & $\cdot$ & $\cdot$ & $\cdot$ & $\cdot$ & $\cdot$ & $\cdot$ & $\cdot$ & $\cdot$ & $\cdot$ & $\cdot$\\
	4 & 0 & 0 & 0 & $\cdot$ & $\cdot$ & $\cdot$ & $\cdot$ & $\cdot$ & $\cdot$ & $\cdot$ & $\cdot$ & $\cdot$ & $\cdot$ & $\cdot$ & $\cdot$\\
	5 & 0 & 0 & 0 & 0 & $\cdot$ & $\cdot$ & $\cdot$ & $\cdot$ & $\cdot$ & $\cdot$ & $\cdot$ & $\cdot$ & $\cdot$ & $\cdot$ & $\cdot$\\
	6 & 0 & 0 & 0 & 0 & 0 & $\cdot$ & $\cdot$ & $\cdot$ & $\cdot$ & $\cdot$ & $\cdot$ & $\cdot$ & $\cdot$ & $\cdot$ & $\cdot$\\
	7 & 0 & 0 & 1 & 1 & 0 & 0 & $\cdot$ & $\cdot$ & $\cdot$ & $\cdot$ & $\cdot$ & $\cdot$ & $\cdot$ & $\cdot$ & $\cdot$\\
	8 & 0 & 0 & 2 & 1 & 2 & 0 & 0 & $\cdot$ & $\cdot$ & $\cdot$ & $\cdot$ & $\cdot$ & $\cdot$ & $\cdot$ & $\cdot$\\
	9 & 0 & 0 & 1 & 7 & 7 & 1 & 0 & 0 & $\cdot$ & $\cdot$ & $\cdot$ & $\cdot$ & $\cdot$ & $\cdot$ & $\cdot$\\
	10 & 0 & 0 & 4 & 8 & 7 & 8 & 4 & 0 & 0 & $\cdot$ & $\cdot$ & $\cdot$ & $\cdot$ & $\cdot$ & $\cdot$\\
	11 & 0 & 0 & 5 & 19 & 33 & 33 & 19 & 5 & 0 & 0 & $\cdot$ & $\cdot$ & $\cdot$ & $\cdot$ & $\cdot$\\
	12 & 0 & 0 & 4 & 14 & 60 & 32 & 60 & 14 & 4 & 0 & 0 & $\cdot$ & $\cdot$ & $\cdot$ & $\cdot$\\
	13 & 0 & 0 & 8 & 39 & 96 & 141 & 141 & 96 & 39 & 8 & 0 & 0 & $\cdot$ & $\cdot$ & $\cdot$\\
	14 & 0 & 0 & 10 & 43 & 148 & 218 & 150 & 218 & 148 & 43 & 10 & 0 & 0 & $\cdot$ & $\cdot$\\
	15 & 0 & 0 & 7 & 70 & 122 & 326 & 582 & 582 & 326 & 122 & 70 & 7 & 0 & 0 & $\cdot$\\
	16 & 0 & 0 & 14 & 55 & 308 & 602 & 1050 & 643 & 1050 & 602 & 308 & 55 & 14 & 0 & 0
	\end{tabular}
\end{center}
\caption{Total genus $g$ of the Riemann surface $\RN$ for small values of $L$ and $N$.}
\label{table g RN}
\end{table}

Lifting closed curves from $\Ch$ with $\pi_{N}^{-1}$, we observe that the ramification points of $\pi_{N}$ are $[\infty,J]=[\infty,A_{\text{out}}J]=\ldots$, whose ramification index is a divisor of $L$, $[0,J]=[0,A_{\text{in}}J]=\ldots$, whose ramification index is a divisor of the least common multiple of $N$ and $L-N$, and $[-1-\rmi\epsilon,J]=[-1+\rmi\epsilon,A_{\text{in}}^{-1}J]=[-1-\rmi\epsilon,A_{\text{out}}A_{\text{in}}^{-1}J]=[-1+\rmi\epsilon,A_{\text{in}}^{-1}A_{\text{out}}A_{\text{in}}^{-1}J]$ with $J$ containing either $1$ or $N+1$ but not both (or equivalently $A_{\text{in}}^{-1}J$ containing either $N$ or $L$ but not both), whose ramification index is equal to $2$.

Using the Riemann-Hurwitz formula (\ref{Riemann Hurwitz}), the ramification indices allow to compute the genus of each connected component of $\RN$. The total genus $g$ of $\RN$, sum of the genus of every connected components, is given in table~\ref{table g RN} for small values of $L$, $N$. Except for the case $N=3$, $L=6$, we observe that $g>0$ as long as $2<N<L-2$.
\end{subsection}

\end{section}

%%%%%%%%%%%%%%%%%%
%%              %%
%%  Section IV  %%
%%              %%
%%%%%%%%%%%%%%%%%%
\begin{section}{Height fluctuations}
\label{section Height fluctuations}
In this section, we obtain exact expressions for multiple point height fluctuations of TASEP in terms of meromorphic differentials on the compact Riemann surface $\RN$. At large $L$, we recover expressions from \cite{P2020.1} for the KPZ fixed point with periodic boundaries, involving the non-compact Riemann surface for half-integer polylogarithms $\R_{\text{KPZ}}$.

\begin{subsection}{Height function}
The evolution in time of the particles of TASEP may be described by occupation numbers $n_{i}(t)\in\{0,1\}$, $n_{1}(t)+\ldots+n_{L}(t)=N$, extended to all $i\in\Z$ by periodicity, $n_{i}(t)=n_{i+L}(t)$. In order to keep track of the total number of particles that have hopped from a site $i$ to the next site $i+1$ (modulo $L$) between time $0$ and time $t$, it is convenient to consider instead TASEP as describing the dynamics of a growing interface, represented by a height function.

Considering an evolution starting from an initial condition $\mathcal{C}_{0}$, we introduce the initial height $H_{i}^{(0)}=\mathrm{H0}_{i}(\mathcal{C}_{0})$, with the function $\mathrm{H0}_{i}$ defined for an arbitrary configuration $\mathcal{C}$ by
\begin{equation}
\label{Hi0}
\mathrm{H0}_{i}(\mathcal{C})=\left\{
\begin{array}{lll}
-\sum_{k=i}^{-1}(\rho-n_{k}) && i<0\\
0 && i=0\\
\sum_{k=1}^{i}(\rho-n_{k}) && i>0
\end{array}\right.\;,
\end{equation}
where $\rho=N/L$ is the average density of particles in the system and $n_{i}$ is the occupation number at site $i$ for the configuration $\mathcal{C}$.
%Considering an evolution starting from an initial condition $\mathcal{C}_{0}$ with occupation numbers $n_{i}^{(0)}=n_{i}(0)$, we define the initial height $H_{i}^{(0)}=\mathrm{H0}_{i}(\mathcal{C}_{0})$ with
%\begin{equation}
%\label{Hi0}
%\mathrm{H0}_{i}(\mathcal{C}_{0})=\left\{
%\begin{array}{lll}
%-\sum_{k=i}^{-1}(\rho-n_{k}^{(0)}) && i<0\\
%0 && i=0\\
%\sum_{k=1}^{i}(\rho-n_{k}^{(0)}) && i>0
%\end{array}\right.\;,
%\end{equation}
%where $\rho=N/L$ is the average density of particles in the system.
The initial height is thus periodic in $i$, $H_{i}^{(0)}=H_{i+L}^{(0)}$, with local increments $H_{i+1}^{(0)}-H_{i}^{(0)}\in\{-(1-\rho),\rho\}$ for all $i$. The dynamics of the TASEP height function $H_{i}(t)$, starting with $H_{i}(0)=H_{i}^{(0)}$, is then defined by increasing $H_{i}(t)$ by $1$ whenever a particle hops from site $i$ to $i+1$ modulo $L$. The evolution preserves periodicity $H_{i}(t)=H_{i+L}(t)$ and local increments $H_{i+1}(t)-H_{i}(t)\in\{-(1-\rho),\rho\}$, and one has at all times
\begin{equation}
\label{Hit[H0t]}
H_{i}(t)=\left\{
\begin{array}{lll}
H_{0}(t)-\sum_{k=i}^{-1}(\rho-n_{k}(t)) && i<0\\
H_{0}(t)+\sum_{k=1}^{i}(\rho-n_{k}(t)) && i>0
\end{array}\right.\;.
\end{equation}
At a given time $t$, height differences $H_{i}(t)-H_{0}(t)$ contain the same information as the configuration of particles, while the quantities $H_{i}(t)-H_{i}(0)$ correspond to the total number of particles that have hopped from sites $i$ to $i+1$ up to time $t$.
\end{subsection}

\begin{subsection}{Deformed Markov operator}
The configuration $\mathcal{C}(t)$ of particles in the system at time $t$ evolves randomly, and the probabilities $\P(\mathcal{C}(t)=\mathcal{C}|\mathcal{C}(0)=\mathcal{C}_{0})$ are solution of the master equation. In the vector space of dimension $|\Omega_{L,N}|={{L}\choose{N}}$ with basis vectors $|\mathcal{C}\rangle$ corresponding to configurations, the probability vector $|P_{t}\rangle_{\mathcal{C}_{0}}=\sum_{\mathcal{C}\in\Omega_{L,N}}\P(\mathcal{C}(t)=\mathcal{C}|\mathcal{C}(0)=\mathcal{C}_{0})\,|\mathcal{C}\rangle$ is equivalently solution of $\partial_{t}|P_{t}\rangle_{\mathcal{C}_{0}}=M|P_{t}\rangle_{\mathcal{C}_{0}}$ with $M$ the Markov matrix of TASEP.

A deformation $M_{i}(\gamma)$ of $M$ allows to compute the joint probability the configuration $\mathcal{C}(t)$ and height $H_{i}(t)$ at a given site $i$ \cite{DL1998.1}: defining
\begin{equation}
\label{F[P]}
\fl\hspace{20mm}
|F_{t,i}\rangle_{\mathcal{C}_{0}}=\sum_{\mathcal{C}\in\Omega_{L,N}}\sum_{U=0}^{\infty}\rme^{\gamma U}\,\P(\mathcal{C}(t)=\mathcal{C},H_{i}(t)=H_{i}(0)+U|\mathcal{C}(0)=\mathcal{C}_{0})\,|\mathcal{C}\rangle\;,
\end{equation}
one has $\partial_{t}|F_{t,i}\rangle_{\mathcal{C}_{0}}=M(\gamma)|F_{t,i}\rangle_{\mathcal{C}_{0}}$, where $\langle\mathcal{C}'|M_{i}(\gamma)|\mathcal{C}\rangle=\rme^{\gamma}$ if $\mathcal{C}'$ is obtained from $\mathcal{C}$ by moving one particle from site $i$ to site $i+1$ and $\langle\mathcal{C}'|M_{i}(\gamma)|\mathcal{C}\rangle=\langle\mathcal{C}'|M|\mathcal{C}\rangle$ otherwise. This implies
\begin{equation}
\label{F[M]}
|F_{t,i}\rangle_{\mathcal{C}_{0}}=\rme^{tM(\gamma)}|\mathcal{C}_{0}\rangle\;.
\end{equation}
Summing over final configurations we obtain the generating function at time $t$ as
\begin{equation}
\label{GF[Mi] one time}
\langle\rme^{\gamma(H_{i}(t)-H_{i}(0))}\rangle_{\mathcal{C}_{0}}=\sum_{\mathcal{C}\in\Omega_{L,N}}\langle\mathcal{C}|\rme^{tM_{i}(\gamma)}|\mathcal{C}_{0}\rangle\;,
\end{equation}
where the averaging on the left is over all TASEP evolutions between time $0$ and time $t$ starting from configuration $\mathcal{C}_{0}$.

Introducing the operator $S_{i}$ defined by
\begin{equation}
\label{Si}
S_{i}|\mathcal{C}\rangle=\Big(\frac{1}{L}\sum_{j=1}^{N}[x_{j}]_{i}\Big)\,|\mathcal{C}\rangle\;,
\end{equation}
with $1\leq x_{1}<\ldots<x_{N}\leq L$ the positions of the particles in the configuration $\mathcal{C}$ and $[x]_{i}$ positions counted from $i$ (i.e. $[i+1]_{i}=1$, $[i+2]_{i}=2$, \ldots, $[L]_{i}=L-i$, $[1]_{i}=L-i+1$, \ldots, $[i-1]_{i}=L-1$, $[i]_{i}=L$), one has the identity
\begin{equation}
\label{Mi[M,Si]}
M_{i}(\gamma)=\rme^{-\gamma S_{i}}\,M(\gamma/L)\,\rme^{\gamma S_{i}}\;,
\end{equation}
where $M(\gamma/L)$, already defined at the beginning of section~\ref{section Bethe ansatz}, corresponds to a deformation spread over all sites: $\langle\mathcal{C}'|M(\gamma/L)|\mathcal{C}\rangle=\rme^{\gamma/L}\langle\mathcal{C}'|M|\mathcal{C}\rangle$ if $\mathcal{C}'\neq\mathcal{C}$ and $\langle\mathcal{C}|M(\gamma/L)|\mathcal{C}\rangle=\langle\mathcal{C}|M|\mathcal{C}\rangle$.
\end{subsection}

\begin{subsection}{Multiple time generating function of the height}
Let $n$ be a positive integer. We fix intermediate times $0=t_{0}<t_{1}<\ldots<t_{n}$ and sites $i_{1},\ldots,i_{n}\in\Z$ defined modulo $L$, and consider the multiple time generating function $\langle\rme^{\sum_{\ell=1}^{n}\gamma_{\ell}(H_{i_{\ell}}(t_{\ell})-H_{i_{\ell}}(0))}\rangle_{\mathcal{C}_{0}}$ for an evolution starting from the configuration $\mathcal{C}_{0}$. Writing $H_{i_{\ell}}(t_{\ell})-H_{i_{\ell}}(0)=\sum_{m=1}^{\ell}(H_{i_{\ell}}(t_{m})-H_{i_{\ell}}(t_{m-1}))$ and introducing the heights $\vec{H}^{(\ell)}=(H_{i}^{(\ell)},i\in\Z)$ at intermediate times $t_{\ell}$, such that $H_{i+1}^{(\ell)}-H_{i}^{(\ell)}\in\{-(1-\rho),\rho\}$, $H_{i+L}^{(\ell)}=H_{i}^{(\ell)}$ and $\vec{H}_{i}^{(\ell+1)}-\vec{H}_{i}^{(\ell)}\in\mathbb{N}$, the Markov property implies
\begin{eqnarray}
&&\fl\hspace{5mm}
\langle\rme^{\sum_{\ell=1}^{n}\gamma_{\ell}(H_{i_{\ell}}(t_{\ell})-H_{i_{\ell}}(0))}\rangle_{\mathcal{C}_{0}}\\
&&\fl\hspace{15mm}
=\sum_{\vec{H}^{(1)},\ldots,\vec{H}^{(n)}}\prod_{\ell=1}^{n}\Big(\rme^{\sum_{m=\ell}^{n}\gamma_{m}(H_{i_{m}}^{(\ell)}-H_{i_{m}}^{(\ell-1)})}\,\P(\vec{H}(t_{\ell})=\vec{H}^{(\ell)}|\vec{H}(t_{\ell-1})=\vec{H}^{(\ell-1)})\Big)\;.\nonumber
\end{eqnarray}
From (\ref{Hit[H0t]}), the sum over the intermediate heights $\vec{H}^{(\ell)}$ can be replaced by a sum over intermediate configurations $\mathcal{C}_{\ell}\in\Omega_{L,N}$ of the particles and height increments $V_{\ell}=H_{0}^{(\ell)}-H_{0}^{(\ell-1)}\in\mathbb{N}$. Using invariance by shifts of time and height, we obtain
\begin{eqnarray}
&&\fl\hspace{5mm}
\langle\rme^{\sum_{\ell=1}^{n}\gamma_{\ell}(H_{i_{\ell}}(t_{\ell})-H_{i_{\ell}}(0))}\rangle_{\mathcal{C}_{0}}
=\sum_{\mathcal{C}_{1},\ldots,\mathcal{C}_{n}\in\Omega_{L,N}}\prod_{\ell=1}^{n}\Big(\sum_{V_{\ell}=0}^{\infty}\rme^{\sum_{m=\ell}^{n}\gamma_{m}(V_{\ell}+\mathrm{H0}_{i_{m}}(\mathcal{C}_{\ell})-\mathrm{H0}_{i_{m}}(\mathcal{C}_{\ell-1}))}\\
&&\fl\hspace{50mm}
\P(\mathcal{C}(t_{\ell}-t_{\ell-1})=\mathcal{C_{\ell}},H_{0}(t_{\ell})=V_{\ell}|\mathcal{C}(0)=\mathcal{C}_{\ell-1},H_{0}(0)=0)\Big)\;,\nonumber
\end{eqnarray}
with $\mathrm{H0}_{i}(\mathcal{C})$ the initial height corresponding to the configuration $\mathcal{C}$ given by (\ref{Hi0}). Comparing with (\ref{F[P]}) and using (\ref{F[M]}), one has
\begin{eqnarray}
&&\fl\hspace{15mm}
\langle\rme^{\sum_{\ell=1}^{n}\gamma_{\ell}(H_{i_{\ell}}(t_{\ell})-H_{i_{\ell}}(0))}\rangle_{\mathcal{C}_{0}}\\
&&\fl\hspace{25mm}
=\sum_{\mathcal{C}\in\Omega_{L,N}}\langle\mathcal{C}|\prod_{\ell=n}^{1}(\rme^{\sum_{m=\ell}^{n}\gamma_{m}\mathrm{H0}_{i_{m}}}\rme^{(t_{\ell}-t_{\ell-1})M_{0}(\sum_{m=\ell}^{n}\gamma_{m})}\rme^{-\sum_{m=\ell}^{n}\gamma_{m}\mathrm{H0}_{i_{m}}})|\mathcal{C}_{0}\rangle\;,\nonumber
\end{eqnarray}
with $\mathrm{H0}_{i}$ the operator defined by $\mathrm{H0}_{i}|\mathcal{C}\rangle=\mathrm{H0}_{i}(\mathcal{C})\,|\mathcal{C}\rangle$. The operator $\mathrm{H0}_{i}$ is related to $S_{i}$ defined in (\ref{Si}) by $\mathrm{H0}_{i}=S_{0}-S_{i}$. Using (\ref{Mi[M,Si]}), we finally obtain after some simplifications
\begin{eqnarray}
\label{GF[M]}
&&\fl\hspace{15mm}
\langle\rme^{\sum_{\ell=1}^{n}\gamma_{\ell}(H_{i_{\ell}}(t_{\ell})-H_{i_{\ell}}(0))}\rangle_{\mathcal{C}_{0}}\\
&&\fl\hspace{25mm}
=\sum_{\mathcal{C}\in\Omega_{L,N}}\langle\mathcal{C}|\Bigg(\prod_{\ell=n}^{1}(\rme^{-\gamma_{\ell}S_{i_{\ell}}}\,\rme^{(t_{\ell}-t_{\ell-1})\,M\big(\!\sum_{m=\ell}^{n}\gamma_{m}/L\big)})\Bigg)\rme^{\sum_{\ell=1}^{n}\gamma_{\ell}S_{i_{\ell}}}|\mathcal{C}_{0}\rangle\;,\nonumber
\end{eqnarray}
which reduces for $n=1$ to the one time generating function (\ref{GF[Mi] one time}) using (\ref{Mi[M,Si]}).
\end{subsection}

\begin{subsection}{Expansion over eigenstates}
The expression (\ref{GF[M]}) can be expanded over left and right Bethe eigenstates $\langle\psi_{r}(\gamma)|$, $|\psi_{r}(\gamma)\rangle$, $r=1,\ldots,|\Omega_{L,N}|$ of the matrices $M(\gamma)$, defined \footnote{For simplicity, we use here notations for the eigenvectors different from the ones in section~\ref{section Bethe ansatz}. In particular, the variable $\gamma$ in the notations $\psi_{r}(\gamma)$, $\psi_{r}^{0}(\gamma)$ indicates that the eigenvector is computed with Bethe roots solution of the Bethe equations (\ref{Bethe equations}) with fugacity $\gamma$.} in terms of Bethe roots in (\ref{psi r}), (\ref{psi l}), with a corresponding eigenvalue $E_{r}(\gamma)$ given by (\ref{E[y]}) for $M(\gamma)$, and an eigenvalue $\rme^{\rmi P_{r}/L}$ given by (\ref{P[y]}) for the translation operator $T$. It is also convenient to introduce the left and right eigenstates $\langle\psi_{r}^{0}(\gamma)|=\langle\psi_{r}(\gamma)|\rme^{L\gamma S_{0}}$ and $|\psi_{r}^{0}(\gamma)\rangle=\rme^{-L\gamma S_{0}}|\psi_{r}(\gamma)\rangle$ of $M_{0}(L\gamma)$, given by (\ref{psi0 r}), (\ref{psi0 l}).

Using the resolution of the identity (\ref{1[psi]}), together with the relations $S_{i}=S_{0}-\mathrm{H0}_{i}$ for the operator $\rme^{\sum_{\ell=1}^{n}\gamma_{\ell}S_{i_{\ell}}}$ and $S_{i}=T^{-i}S_{0}T^{i}$ for the operator $\rme^{-\gamma_{\ell}S_{i_{\ell}}}$, and finally the translation invariance of the vector $\sum_{\mathcal{C}\in\Omega_{L,N}}\langle\mathcal{C}|$, we obtain from (\ref{GF[M]})
\begin{eqnarray}
\label{GF[psi0]}
&&\fl\hspace{5mm}
\Big\langle\rme^{\sum_{\ell=1}^{n}\gamma_{\ell}\,H_{i_{\ell}}(t_{\ell})}\Big\rangle_{\mathcal{C}_{0}}
=\sum_{r_{1},\ldots,r_{n}=1}^{|\Omega_{L,N}|}\left(\prod_{\ell=1}^{n}\frac{\rme^{(t_{\ell}-t_{\ell-1})E_{r_{\ell}}\big(\sum\limits_{m=\ell}^{n}\gamma_{m}/L\big)-\rmi(i_{\ell}-i_{\ell-1})P_{r_{\ell}}/L}}{\Big\langle\psi_{r_{\ell}}^{0}\Big(\sum\limits_{m=\ell}^{n}\frac{\gamma_{m}}{L}\Big)\Big|\psi_{r_{\ell}}^{0}\Big(\sum\limits_{m=\ell}^{n}\frac{\gamma_{m}}{L}\Big)\Big\rangle}\right)\nonumber\\
&&\fl\hspace{55mm}
\times\Bigg(\sum_{\mathcal{C}\in\Omega_{L,N}}\Big\langle\mathcal{C}\Big|\psi_{r_{n}}^{0}\Big(\frac{\gamma_{n}}{L}\Big)\Big\rangle\Bigg)\Big\langle\psi_{r_{1}}^{0}\Big(\sum\limits_{m=1}^{n}\frac{\gamma_{m}}{L}\Big)\Big|\mathcal{C}_{0}\Big\rangle\\
&&\fl\hspace{55mm}
\times\Bigg(\prod_{\ell=1}^{n-1}\Big\langle\psi_{r_{\ell+1}}^{0}\Big(\sum\limits_{m=\ell+1}^{n}\frac{\gamma_{m}}{L}\Big)\Big|\psi_{r_{\ell}}^{0}\Big(\sum\limits_{m=\ell}^{n}\frac{\gamma_{m}}{L}\Big)\Big\rangle\Bigg)\;,\nonumber
\end{eqnarray}
with the convention $t_{0}=0$, $i_{0}=0$.
\end{subsection}

\begin{subsection}{Bethe ansatz formula for the generating function}
Using the results of section~\ref{section Bethe ansatz} for eigenvalues and eigenvectors in terms of Bethe roots, the expression (\ref{GF[psi0]}) for the generating function can be rewritten as
\begin{eqnarray}
\label{GF[y]}
&&\fl\hspace{2mm}
\Big\langle\rme^{\sum_{\ell=1}^{n}\gamma_{\ell}\,H_{i_{\ell}}(t_{\ell})}\Big\rangle_{\mathcal{C}_{0}}
=\Big(\prod_{\ell=1}^{n}\sum_{\vec{y}^{(\ell)}}\Big)
\frac{\det\Big((y_{j}^{(1)})^{k-1}(1-y_{j}^{(1)})^{-x_{k}^{(0)}}\Big)_{j,k\in\[1,N\]}}{\prod_{j=1}^{N}\prod_{k=j+1}^{N}(y_{j}^{(1)}-y_{k}^{(1)})}
\,\frac{1}{\prod_{j=1}^{N}(y_{j}^{(n)})^{N}}
\nonumber\\
&&\fl\hspace{14mm}
\times\Bigg(\prod_{\ell=1}^{n-1}\frac{(-1)^{\frac{N(N-1)}{2}}\Big(1-\frac{\rme^{\sum_{m=\ell+1}^{n}\gamma_{m}}\prod_{j=1}^{N}y_{j}^{(\ell+1)}}{\rme^{\sum_{m=\ell}^{n}\gamma_{m}}\prod_{j=1}^{N}y_{j}^{(\ell)}}\Big)^{N-1}}{\prod_{j=1}^{N}\prod_{k=1}^{N}(y_{j}^{(\ell)}-y_{k}^{(\ell+1)})}\Bigg)
\Bigg(\prod_{\ell=1}^{n}\prod_{j=1}^{N}\prod_{k=j+1}^{N}(y_{j}^{(\ell)}-y_{k}^{(\ell)})^{2}\Bigg)\nonumber\\
&&\fl\hspace{5mm}
\times\prod_{\ell=1}^{n}\frac{(1-\rme^{-\gamma_{\ell}})\,\rme^{\frac{Ni_{\ell}\gamma_{\ell}}{L}}\,(\prod_{j=1}^{N}y_{j}^{(\ell)}(1-y_{j}^{(\ell)})^{1+i_{\ell}-i_{\ell-1}})\,\rme^{(t_{\ell}-t_{\ell-1})\sum\limits_{j=1}^{N}\frac{y_{j}^{(\ell)}}{1-y_{j}^{(\ell)}}}}{\Big(\frac{L}{N}\sum_{j=1}^{N}\frac{y_{j}^{(\ell)}}{N+(L-N)y_{j}^{(\ell)}}\Big)\prod_{j=1}^{N}(N+(L-N)y_{j}^{(\ell)})}\;,
\end{eqnarray}
where the summation is over all $|\Omega_{L,N}|$ admissible solutions $\vec{y}^{(\ell)}=(y_{1}^{(\ell)},\ldots,y_{N}^{(\ell)})$ of the Bethe equations (\ref{Bethe equations}) with fugacity $\sum_{m=\ell}^{n}\gamma_{m}/L$.
\end{subsection}

\begin{subsection}{Probability of the height}
The height difference $H_{i}(t)-H_{i}(0)$ is a non-negative integer, and the joint probability of the height can thus be computed from the generating function above by taking residues, as
\begin{equation}
\fl\hspace{5mm}
\P(H_{i_{\ell}}(t_{\ell})=H_{i_{\ell}}^{(0)}+U_{\ell},\ell=1,\ldots,n|\mathcal{C}_{0})
=\oint\Big(\prod_{\ell=1}^{n}\frac{\rmd g_{\ell}}{g_{\ell}^{1+H_{i_{\ell}}^{(0)}+U_{\ell}}}\Big)\,\Big\langle\prod_{\ell=1}^{n}g_{\ell}^{H_{i_{\ell}}(t_{\ell})}\Big\rangle_{\mathcal{C}_{0}}\;.
\end{equation}
The contours of integration encircle $0$ once in the positive direction. Inserting (\ref{GF[y]}), we obtain an expression for the joint probability in terms of sums over solutions $\vec{y}^{(\ell)}$, $\ell=1,\ldots,n$ of the Bethe equations for fugacities $\sum_{m=\ell}^{n}\gamma_{m}/L$ with $g_{m}=\rme^{\gamma_{m}}$. In order to rewrite the probability directly in terms of differentials on the Riemann surface $\RN$, it is useful in view of (\ref{C[y]}) to make the change of variables
\begin{equation}
g_{\ell}\to C_{\ell}=\frac{(\prod_{m=\ell}^{n}g_{m})\prod_{j=1}^{N}y_{j}^{(\ell)}}{\rho^{N}(1-\rho)^{L-N}}\;.
\end{equation}
The Jacobian
\begin{equation}
\fl\hspace{15mm}
\det(\partial_{C_{\ell}}g_{m})_{l,m=1,\ldots,n}=\frac{\rho^{N}(1-\rho)^{L-N}C_{1}}{\prod_{j=1}^{N}y_{j}^{(1)}}\,\prod_{\ell=1}^{n}\Big(\frac{1}{C_{\ell}}\,\frac{L}{N}\sum_{j=1}^{N}\frac{y_{j}^{(\ell)}}{N+(L-N)y_{j}^{(\ell)}}\Big)
\end{equation}
cancels some factors in (\ref{GF[y]}). Introducing the functions $y_{j}(C)$, $j\in\[1,L\]$ from section~\ref{section Riemann surfaces}, the summation over Bethe roots $\vec{y}^{(\ell)}$ can be replaced by a sum over sets $\{j_{1}^{(\ell)},\ldots,j_{N}^{(\ell)}\}\subset\[1,L\]$ indexing the sheets of $\RN$ corresponding to the covering map $\pi_{N}$, and we obtain
\begin{eqnarray}
\label{P[H=U]}
&&\fl\hspace{5mm}
\P(H_{i_{\ell}}(t_{\ell})=H_{i_{\ell}}^{(0)}+U_{\ell},\ell=1,\ldots,n|\mathcal{C}_{0})
=\oint\Big(\prod_{\ell=1}^{n}\frac{\rmd C_{\ell}}{2\rmi\pi\,C_{\ell}}\Big)
\Big(\prod_{\ell=1}^{n}\sum_{1\leq j_{1}^{(\ell)}<\ldots<j_{N}^{(\ell)}\leq L}\Big)\nonumber\\
&&\fl\hspace{15mm}
	\frac{\det\Big((y_{\lambda}^{(1)})^{k-1}(1-y_{\lambda}^{(1)})^{-x_{k}^{(0)}}\Big)_{k,\lambda\in\[1,N\]}}{\prod_{\kappa=1}^{N}\prod_{\lambda=\kappa+1}^{N}(y_{\kappa}^{(1)}-y_{\lambda}^{(1)})}
	\;\frac{1-\frac{\prod_{\kappa=1}^{N}y_{\kappa}^{(n)}}{\rho^{N}(1-\rho)^{L-N}C_{n}}}{\prod_{\kappa=1}^{N}(y_{\kappa}^{(n)})^{N}}\\
&&\fl\hspace{10mm}
\times\prod_{\ell=1}^{n}
\Bigg(
	\Big(\prod_{\kappa=1}^{N}\frac{y_{\kappa}^{(\ell)}(1-y_{\kappa}^{(\ell)})}{N+(L-N)y_{\kappa}^{(\ell)}}\Big)
	\Big(\prod_{\kappa=1}^{N}\prod_{\lambda=\kappa+1}^{N}(y_{\kappa}^{(\ell)}-y_{\lambda}^{(\ell)})^{2}\Big)
	\Big(\prod_{\kappa=1}^{N}(1-y_{\kappa}^{(\ell)})\Big)^{i_{\ell}-i_{\ell-1}}\nonumber\\
&&
	\Big(\frac{\prod_{\kappa=1}^{N}y_{\kappa}^{(\ell)}}{\rho^{N}(1-\rho)^{L-N}C_{\ell}}\Big)^{(H_{i_{\ell}}^{(0)}+U_{\ell}-Ni_{\ell}/L)-(H_{i_{\ell-1}}^{(0)}+U_{\ell-1}-Ni_{\ell-1}/L)}
	\,\rme^{(t_{\ell}-t_{\ell-1})\sum\limits_{\kappa=1}^{N}\frac{y_{\kappa}^{(\ell)}}{1-y_{\kappa}^{(\ell)}}}
\Bigg)\nonumber\\
&&\fl\hspace{10mm}
\times\prod_{\ell=1}^{n-1}\frac{(-1)^{\frac{N(N-1)}{2}}\Big(1-\frac{C_{\ell+1}\prod_{\kappa=1}^{N}y_{\kappa}^{(\ell)}}{C_{\ell}\prod_{\kappa=1}^{N}y_{\kappa}^{(\ell+1)}}\Big)\,\Big(1-\frac{C_{\ell+1}}{C_{\ell}}\Big)^{N-1}}{\prod_{\kappa=1}^{N}\prod_{\lambda=1}^{N}(y_{\kappa}^{(\ell)}-y_{\lambda}^{(\ell+1)})}\;,\nonumber
\end{eqnarray}
with the notation $y_{\kappa}^{(\ell)}=y_{j_{\kappa}^{(\ell)}}(C_{\ell})$.

We emphasize that the summand starting at line 2 in (\ref{P[H=U]}) is not meromorphic in $\C$, since the functions $y_{j}$ have branch cuts. For any $\ell=1,\ldots,n$, we observe however that the summand is a symmetric function of the $y_{\kappa}^{(\ell)}$, $\kappa\in\[1,N\]$, and can thus be interpreted as the evaluation at the point $[C_{\ell},\{j_{1}^{(\ell)},\ldots,j_{N}^{(\ell)}\}]$ of a function meromorphic on $\RN$ (except for essential singularities at points $[\infty,J]$, see below). The sum over all sets $\{j_{1}^{(\ell)},\ldots,j_{N}^{(\ell)}\}\subset\[1,L\]$ labelling the sheets of $\RN$ with respect to the covering map $\pi_{N}$, called the \emph{trace} over $\pi_{N}$ of the function on $\RN$, see e.g. \cite{S1957.1}, is then meromorphic in $\C$, with an essential singularity at infinity.

We discuss in the rest of this section the pole structure of the summand in (\ref{P[H=U]}). From the results of section~\ref{section Riemann surfaces}, factors $y_{j}(C)$ may only have poles and zeroes at the points $[0,J]$ while factors $1-y_{j}(C)$ may only have poles at the points $[0,J]$ and zeroes at the points $[\infty,J]$. We recall that since the points $[0,J]$ and $[\infty,J]$ are ramification points of $\pi_{N}$, the small and large $C$ expansions of symmetric functions of $N$ distinct $y_{j}(C)$ may contain fractional powers of $C$. Such functions are however perfectly meromorphic on $\RN$ after an appropriate choice of local parameter around these points.

The ratio $\det((y_{\lambda}^{(1)})^{k-1}(1-y_{\lambda}^{(1)})^{-x_{k}^{(0)}})_{k,\lambda\in\[1,N\]}/\prod_{\kappa=1}^{N}\prod_{\lambda=\kappa+1}^{N}(y_{\kappa}^{(1)}-y_{\lambda}^{(1)})$ is a symmetric polynomial in the $(1-y_{\lambda}^{(1)})^{-1}$, whose only poles are then of the form $[\infty,J]$. Thus, the second line of (\ref{P[H=U]}) may only have poles in the variables $C_{1}$ and $C_{n}$ at the points $[0,J]$, $[\infty,J]$. Similarly, in the third line of (\ref{P[H=U]}), except for the denominator $N+(L-N)y_{\kappa}^{(\ell)}$ discussed separately at the end of this section, all the factors may only have poles in the variables $C_{\ell}$ at the points $[0,J]$, $[\infty,J]$. The same is true for the fourth line of (\ref{P[H=U]}), with the exponential contributing an additional essential singularity, i.e. a pole of infinite order, at the points $[\infty,J]$.

In the last line of (\ref{P[H=U]}), in addition to the usual poles in the variables $C_{\ell}$ at the points $[0,J]$, the denominator vanishes when $C_{\ell}=C_{\ell+1}$ and $j_{\kappa}^{(\ell)}=j_{\lambda}^{(\ell+1)}$. If the sets $\{j_{\kappa}^{(\ell)}\}$ and $\{j_{\lambda}^{(\ell+1)}\}$ are not identical, the apparent pole at $C_{\ell}=C_{\ell+1}$ coming from the denominator, of order at most $N-1$, is compensated by the factor $(1-C_{\ell+1}/C_{\ell})^{N-1}$ in the numerator. If $\{j_{\kappa}^{(\ell)}\}=\{j_{\lambda}^{(\ell+1)}\}$, the apparent pole at $C_{\ell}=C_{\ell+1}$ from the denominator is of order $N$, and is again fully compensated by the factors $(1-C_{\ell+1}/C_{\ell})^{N-1}$ and $1-\frac{C_{\ell+1}\prod_{\kappa=1}^{N}y_{\kappa}^{(\ell)}}{C_{\ell}\prod_{\kappa=1}^{N}y_{\kappa}^{(\ell+1)}}$ in the numerator. The last line of (\ref{P[H=U]}) has thus only poles at the points $[0,J]$ in the variables $C_{\ell}$, and stays finite at $C_{\ell}=C_{\ell+1}$.

We consider now poles at $C_{\ell}=-1$, contributed only by the factors in the third line of (\ref{P[H=U]}) corresponding to the function $\Pi$ on $\RN$ defined by $\Pi([C,J])=\prod_{j<k\in J}(y_{j}(C)-y_{k}(C))^{2}/\prod_{j\in J}(N+(L-N)y_{j}(C))$. From the results of section~\ref{section Riemann surfaces}, the factor $N+(L-N)y_{j}(C)$ vanishes as $\sqrt{C+1}$ when either $j\in\{1,N+1\}$ and $C\to-1$ with $\Im~C<0$ or $j\in\{N,L\}$ and $C\to-1$ with $\Im~C>0$. There are three cases to consider for the behaviour of the function $\Pi$ at the point $p=[-1-\rmi\epsilon,J]$ (respectively $p=[-1+\rmi\epsilon,J]$), $0<\epsilon\to0$ depending on the set $I=J\cap\{1,N+1\}$ (resp. $I=J\cap\{N,L\}$). If $I$ is empty, then $\Pi$ stays finite at the point $p$. If $I$ contains a single element, then $\Pi([C,J])\sim(C+1)^{-1/2}$. Furthermore, the point $p$ is a ramification point of $\pi_{N}$ with ramification index $2$ in this case, with local parameter $\sqrt{C+1}$ around $p$, and the function $\Pi$ has thus a simple pole at $p$. Finally, if $I$ contains two elements, $\prod_{j\in J}(N+(L-N)y_{j}(C))\sim C+1$ is compensated by $\prod_{j<k\in J}(y_{j}(C)-y_{k}(C))^{2}\sim C+1$, and the function $\Pi$ is again finite at $p$, which is furthermore not a ramification point of $\pi_{N}$ in that case. We conclude that the function $\Pi$, and thus all the summand between the second and last line of (\ref{P[H=U]}), has simple poles at the points $[-1,J]\in\RN$ only if they are also ramification points of $\pi_{N}$.

We consider finally the differential form $\Omega$, defined away from ramification points of $\pi_{N}$ by $\Omega([C,J])=\Pi([C,J])\,\rmd C$. Around a ramification point of $\pi_{N}$ of the form $p=[-1\pm\rmi\epsilon,J]$, an appropriate choice of local parameter is $B=\sqrt{C+1}$, and we observe that the zero of $\rmd C=2B\,\rmd B$ compensates the pole of the function $\Pi$ at $p$. The differential form $\Omega$ is thus analytic at $p$.

In conclusion, we have shown that the integrand in (\ref{P[H=U]}), interpreted for any variable $C_{\ell}$ as a differential living on $\RN$, only has poles at the points $[0,J]$ and $[\infty,J]$ and is regular everywhere else. The trace obtained by summing over all sheets indexed by sets $\{j_{1}^{(\ell)},\ldots,j_{N}^{(\ell)}\}\subset\[1,L\]$ is thus holomorphic in $\C^{*}$, with a multiple pole at $C_{\ell}=0$ and a pole of infinite order at $C_{\ell}=\infty$. The contours of integration in (\ref{P[H=U]}) are thus only required to encircle $0$ once in the positive direction, and can be moved freely beyond that.
\end{subsection}

\begin{subsection}{Cumulative distribution of the height}
The cumulative distribution of the height
\begin{eqnarray}
&&
\P(H_{i_{\ell}}(t_{\ell})\geq H_{i_{\ell}}^{(0)}+U_{\ell},\ell=1,\ldots,n|\mathcal{C}_{0})\\
&&\hspace{35mm}
=\Big(\prod_{\ell=1}^{n}\sum_{V_{\ell}=U_{\ell}}^{\infty}\Big)\P(H_{i_{\ell}}(t_{\ell})=H_{i_{\ell}}^{(0)}+V_{\ell},\ell=1,\ldots,n|\mathcal{C}_{0})\nonumber
\end{eqnarray}
follows easily from (\ref{P[H=U]}). In order to perform the summation over the $V_{\ell}$ inside of the integrals, the additional constraint
\begin{equation}
\Bigg|\frac{\prod_{\kappa=1}^{N}y_{j_{\kappa}^{(1)}}(C_{1})}{\rho^{N}(1-\rho)^{L-N}C_{1}}\Bigg|<\ldots<\Bigg|\frac{\prod_{\kappa=1}^{N}y_{j_{\kappa}^{(n)}}(C_{n})}{\rho^{N}(1-\rho)^{L-N}C_{n}}\Bigg|<1
\end{equation}
is needed for all possible choices of the integers $j_{\kappa}^{(\ell)}$. Since the Bethe root functions $y_{j}(C)\to1$ when $|C|\to\infty$, the constraint implies the ordering of the paths of integration
\begin{equation}
\label{ordering paths}
|C_{n}|<\ldots<|C_{1}|
\end{equation}
in the region where the $|C_{\ell}|$ are large. The pole structure discussed at the end of this section shows that the ordering (\ref{ordering paths}) is in fact necessary and sufficient even when the $|C_{\ell}|$ are not large. We obtain
\begin{eqnarray}
\label{P[H>=U]}
&&\fl\hspace{5mm}
\P(H_{i_{\ell}}(t_{\ell})\geq H_{i_{\ell}}^{(0)}+U_{\ell},\ell=1,\ldots,n|\mathcal{C}_{0})
=\oint_{|C_{n}|<\ldots<|C_{1}|}\Big(\prod_{\ell=1}^{n}\frac{\rmd C_{\ell}}{2\rmi\pi\,C_{\ell}}\Big)\\
&&\fl\hspace{7mm}
\Big(\prod_{\ell=1}^{n}\sum_{1\leq j_{1}^{(\ell)}<\ldots<j_{N}^{(\ell)}\leq L}\Big)\frac{\det\Big((y_{\lambda}^{(1)})^{k-1}(1-y_{\lambda}^{(1)})^{-x_{k}^{(0)}}\Big)_{k,\lambda\in\[1,N\]}}{\prod_{\kappa=1}^{N}\prod_{\lambda=\kappa+1}^{N}(y_{\kappa}^{(1)}-y_{\lambda}^{(1)})}
\;\frac{1}{\prod_{\kappa=1}^{N}(y_{\kappa}^{(n)})^{N}}\nonumber\\
&&\fl\hspace{10mm}
\times\prod_{\ell=1}^{n}
\Bigg(
\Big(\prod_{\kappa=1}^{N}\frac{y_{\kappa}^{(\ell)}(1-y_{\kappa}^{(\ell)})}{N+(L-N)y_{\kappa}^{(\ell)}}\Big)
\Big(\prod_{\kappa=1}^{N}\prod_{\lambda=\kappa+1}^{N}(y_{\kappa}^{(\ell)}-y_{\lambda}^{(\ell)})^{2}\Big)
\Big(\prod_{\kappa=1}^{N}(1-y_{\kappa}^{(\ell)})\Big)^{i_{\ell}-i_{\ell-1}}\nonumber\\
&&
\Big(\frac{\prod_{\kappa=1}^{N}y_{\kappa}^{(\ell)}}{\rho^{N}(1-\rho)^{L-N}C_{\ell}}\Big)^{(H_{i_{\ell}}^{(0)}+U_{\ell}-Ni_{\ell}/L)-(H_{i_{\ell-1}}^{(0)}+U_{\ell-1}-Ni_{\ell-1}/L)}
\,\rme^{(t_{\ell}-t_{\ell-1})\sum\limits_{\kappa=1}^{N}\frac{y_{\kappa}^{(\ell)}}{1-y_{\kappa}^{(\ell)}}}
\Bigg)\nonumber\\
&&\fl\hspace{10mm}
\times\prod_{\ell=1}^{n-1}\frac{(-1)^{\frac{N(N-1)}{2}}\,\Big(1-\frac{C_{\ell+1}}{C_{\ell}}\Big)^{N-1}}{\prod_{\kappa=1}^{N}\prod_{\lambda=1}^{N}(y_{\kappa}^{(\ell)}-y_{\lambda}^{(\ell+1)})}\;,\nonumber
\end{eqnarray}
with the same notations $y_{\kappa}^{(\ell)}=y_{j_{\kappa}^{(\ell)}}(C_{\ell})$ as before. The identity (\ref{P[H>=U]}) is equivalent to the expression (3.6) from \cite{BL2019.1} after some rewriting, with $z_{\ell}^{L}=\rho^{N}(1-\rho)^{L-N}C_{\ell}$.

As in the previous section, the summand in (\ref{P[H>=U]}) can be interpreted as a function of the $n$ variables $p_{\ell}=[C_{\ell},\{j_{1}^{(\ell)},\ldots,j_{N}^{(\ell)}\}]\in\RN$, $\ell=1,\ldots,n$, meromorphic (except for isolated essential singularities) in each variable with poles of finite order at points $[0,J]$ and poles of infinite order at points $[\infty,J]$. Unlike in (\ref{P[H=U]}), however, this function has the additional poles $p_{\ell}=p_{\ell+1}$, coming from the denominator $\prod_{\kappa=1}^{N}\prod_{\lambda=1}^{N}(y_{\kappa}^{(\ell)}-y_{\lambda}^{(\ell+1)})$ with $\{j_{1}^{(\ell)},\ldots,j_{N}^{(\ell)}\}=\{j_{1}^{(\ell+1)},\ldots,j_{N}^{(\ell+1)}\}$, which is no longer compensated by a factor $1-\frac{C_{\ell+1}}{C_{\ell}}\prod_{\kappa=1}^{N}y_{\kappa}^{(\ell)}/y_{\kappa}^{(\ell+1)}$ in the numerator. This implies that the contours of integration can not be moved freely, but are constrained by (\ref{ordering paths}) after taking the trace over all sheets.
\end{subsection}

\begin{subsection}{Abelian integrals when $L$ and $N$ are co-prime}
We introduce meromorphic functions $\mu_{1}$, $\mu_{2}$ and $\eta$ on the Riemann surface $\RN$ by
\begin{eqnarray}
\label{mu1}
&& \mu_{1}([C,J])=-1+\sum_{j\in J}\frac{1}{N+(L-N)y_{j}(C)}\\
\label{mu2}
&& \mu_{2}([C,J])=-\frac{1}{N}+\sum_{j\in J}\frac{1}{(N+(L-N)y_{j}(C))^{2}}\\
\label{eta}
&& \eta([C,J])=-N+\sum_{j\in J}\frac{1}{1-y_{j}(C)}\;.
\end{eqnarray}
These functions behave as $\mathcal{O}(C)$ when $C\to0$ in the principal sheet $J=\[1,N\]$. Additionally, for any meromorphic function $f$ on $\RN$, we use the notation $\omega_{f}$ for the meromorphic differential defined away from ramification points of $\pi_{N}$ by $\omega_{f}([C,J])=f([C,J])\rmd C/C$.

When $L$ and $N$ are co-prime, the Riemann surface $\RN$ has a single connected component, and any point $[C,J]$ can be reached from the point
\begin{equation}
\label{O}
O=[0,\[1,N\]]\in\RN\;,
\end{equation}
which is never a ramification point of $\pi_{N}$. From the expression (\ref{yj'}) of $y_{j}'(C)$, and the small $C$ behaviour (\ref{yj C0}) of $y_{j}(C)$ to fix the constant of integration, we obtain expressions involving Abelian integrals (i.e. integrals of meromorphic differentials on a Riemann surface) for the sum
\begin{equation}
\label{identity sum}
\sum_{j\in J}\frac{y_{j}(C)}{1-y_{j}(C)}=\frac{1}{L}\int_{O}^{[C,J]}\omega_{\eta}-\frac{N}{L}\int_{O}^{[C,J]}\omega_{\mu_{1}}\;,
\end{equation}
the products
\begin{eqnarray}
&& \prod_{j\in J}^{N}\frac{y_{j}(C)\,(1-y_{j}(C))}{N+(L-N)y_{j}(C)}=\frac{(1-\rho)^{L-N}C}{L^{N}}\,\exp\Big(\frac{NL}{L-N}\int_{O}^{[C,J]}\omega_{\mu_{2}}\Big)\\
&& \prod_{j\in J}(1-y_{j}(C))=\exp\Big(\frac{N}{L-N}\int_{O}^{[C,J]}\omega_{\mu_{1}}\Big)\\
&& \prod_{j\in J}\frac{y_{j}(C)}{\rho^{N}(1-\rho)^{L-N}C}=\exp\Big(\frac{L}{L-N}\int_{O}^{[C,J]}\omega_{\mu_{1}}\Big)
\end{eqnarray}
and the double products
\begin{eqnarray}
&& \prod_{{j,k\in J}\atop{j<k}}(y_{j}(C)-y_{k}(C))^{2}=(-1)^{\frac{N(N-1)}{2}}L^{N}\rho^{N^{2}}(1-\rho)^{(N-1)(L-N)}C^{N-1}\\
&&\hspace{50mm} \times\exp\Big(\frac{NL}{L-N}\int_{O}^{[C,J]}\omega_{\mu_{1}^{2}+2\mu_{1}-\mu_{2}}\Big)\nonumber
\end{eqnarray}
and
\begin{eqnarray}
\label{identity prod prod 12}
&&
\prod_{j_{1}\in J_{1}}\prod_{j_{2}\in J_{2}}(y_{j_{1}}(C_{1})-y_{j_{2}}(C_{2}))=\rho^{N^{2}}(1-\rho)^{N(L-N)}(C_{1}-C_{2})^{N}\\
&&\hspace{35mm}
\times\exp\Bigg(\frac{NL}{L-N}\Big(\int_{\gamma}\frac{\rmd B}{B}\,\mathcal{A}(\mu_{1}([C_{1}B,\cdot])\mu_{1}([C_{2}B,\cdot]))\nonumber\\
&&\hspace{72mm}
+\int_{O}^{[C_{1},J_{1}]}\omega_{\mu_{1}}+\int_{O}^{[C_{2},J_{2}]}\omega_{\mu_{1}}\Big)\Bigg)\;.\nonumber
\end{eqnarray}
The path $\gamma\subset\C\setminus\{0,-1\}$ in (\ref{identity prod prod 12}) is chosen so that $\gamma$ lifts for the couple $([C_{1}B,\cdot],[C_{2}B,\cdot])\in\RN\times\RN$ to a path $\Gamma$ from $(O,O)$ to $([C_{1},J_{1}],[C_{2},J_{2}])$, and $\mathcal{A}(\mu_{1}([C_{1}B,\cdot])\mu_{1}([C_{2}B,\cdot]))$ is understood as the analytic continuation of $\mu_{1}([C_{1}B,\cdot])\mu_{1}([C_{2}B,\cdot])$ along $\Gamma$.

Using the integral formulas above, we finally obtain from (\ref{P[H>=U]}) our main result for the joint statistics of the TASEP height at $n$ distinct times $0<t_{1}<\ldots<t_{n}$, valid when the system size $L$ and the number of particles $N$ are co-prime \footnote{When $L$ and $N$ are not co-prime, and additional summation over the connected components of $\RN$ is needed, and origin points analogue to $O$ must be chosen for each connected component. The definitions of the functions $\mu_{1}$ and $\eta$ must also be altered accordingly in order to make the integrals convergent.}:
\begin{eqnarray}
\label{P[H>=U] exp}
&&\fl\hspace{1mm}
\P(H_{i_{\ell}}(t_{\ell})\geq H_{i_{\ell}}^{(0)}+U_{\ell},\ell=1,\ldots,n|\mathcal{C}_{0})
=\oint_{|C_{n}|<\ldots<|C_{1}|}
\!\!\frac{\rmd C_{1}\ldots\rmd C_{n}}{(2\rmi\pi)^{n}\,C_{n}}\;
\Big(\prod_{\ell=1}^{n}\sum_{J_{\ell}\subset\[1,L\],\;|J_{\ell}|=N}\Big)\nonumber\\
&&\fl\hspace{1mm}
\Theta_{\vec{x_{0}}}([C_{1},J_{1}])\;
\frac{
\prod_{\ell=1}^{n}\rme^{\int_{O}^{[C_{\ell},J_{\ell}]}\big(\frac{NL}{L-N}\,\omega_{\mu_{1}^{2}}+\big((H_{i_{\ell}}^{(0)}+U_{\ell})-(H_{i_{\ell-1}}^{(0)}+U_{\ell-1})\big)\frac{L\,\omega_{\mu_{1}}}{L-N}+(t_{\ell}-t_{\ell-1})\big(\frac{\omega_{\eta}}{L}-\frac{N\,\omega_{\mu_{1}}}{L}\big)\big)}
}
{
\prod_{\ell=1}^{n-1}\Big((C_{\ell}-C_{\ell+1})\;\rme^{\frac{NL}{L-N}\int_{\gamma}\frac{\rmd B}{B}\,\mathcal{A}(\mu_{1}([C_{\ell}B,\cdot])\mu_{1}([C_{\ell+1}B,\cdot]))}\Big)
}\,,
\end{eqnarray}
with initial heights $H_{i}^{(0)}$ defined in (\ref{Hi0}), integer height differences $U_{\ell}$, and the conventions $t_{0}=i_{0}=U_{0}=0$. As above, the meromorphic differentials are defined by $\omega_{f}([C,J])=f([C,J])\,\rmd C/C$ away from branch points of $\pi_{N}$, the functions $\mu_{1}$ and $\eta$ on $\RN$ are defined by (\ref{mu1}) and (\ref{eta}), the point $O=[0,\[1,N\]]$ belongs to the principal sheet of $\RN$, the path $\gamma\subset\C\setminus\{0,-1\}$ from $0$ to $1$ is as in (\ref{identity prod prod 12}), and the symbol $\mathcal{A}$ means analytic continuation on $\RN\times\RN$ along $\gamma$. The summation over sheets $J_{\ell}$ corresponds to tracing over the covering map $\pi_{N}$ from $\RN$ to the Riemann sphere $\Ch$. The initial positions $x_{k}^{(0)}\in\[1,L\]$ of the particles enter only through the ratio
\begin{equation}
\fl\hspace{10mm}
\Theta_{x_{1}^{(0)},\ldots,x_{N}^{(0)}}([C,\{j_{1},\ldots,j_{N}\}])=\frac{\det\Big((y_{j_{\lambda}}(C))^{k-1}(1-y_{j_{\lambda}}(C))^{L-x_{k}^{(0)}}\Big)_{k,\lambda\in\[1,N\]}}{\prod_{\kappa=1}^{N}\prod_{\lambda=\kappa+1}^{N}(y_{j_{\lambda}}(C)-y_{j_{\kappa}}(C))}\;.
\end{equation}
For domain wall initial condition $x_{k}^{(0)}=k+i$, $0\leq i\leq L-N$ one has in particular
\begin{equation}
\fl\hspace{20mm}
\Theta_{\text{dw}}([C,J])
=\prod_{j\in J}(1-y_{j}(C))^{L-N-i}
=\exp\Big(\frac{N(L-N-i)}{L-N}\int_{O}^{[C,J]}\omega_{\mu_{1}}\Big)\;.
\end{equation}
For the stationary initial condition, which consists in summing over all possible choices of initial positions with the same weight $1/|\Omega_{L,N}|$, the identity $\sum_{1\leq x_{1}<\ldots<x_{N}\leq L}\Theta_{x_{1},\ldots,x_{N}}([C,J])=(1-\sum_{j\in J}(1-y_{j}(C))^{L}\prod_{k\in J\setminus\{j\}}\frac{y_{k}(C)}{y_{k}(C)-y_{j}(C)})/\prod_{j\in J}y_{j}(C)$ gives, after using $P(y_{j}(C),C)=0$ and computing explicitly the sum over $j\in J$, the result
\begin{equation}
\fl\hspace{2mm}
\Theta_{\text{stat}}([C,J])
=\frac{1}{|\Omega_{L,N}|}\Big(\frac{1}{\prod_{j\in J}y_{j}(C)}-\frac{1}{\rho^{N}(1-\rho)^{L-N}C}\Big)=\frac{-1+\rme^{-\frac{L}{L-N}\int_{O}^{[C,J]}\omega_{\mu_{1}}}}{|\Omega_{L,N}|\,\rho^{N}(1-\rho)^{L-N}C}\;.
\end{equation}
More generally, the symmetric Grothendieck polynomial $\Theta_{x_{1}^{(0)},\ldots,x_{N}^{(0)}}([C,J])$ may be written as a sum over line ensembles corresponding to interlacing sequences of positions \cite{MS2014.2}, and each term of the sum is proportional to the exponential of an integral over $\omega_{\tilde{\mu}_{1}}$, with $\tilde{\mu}_{1}$ an inhomogeneous version of $\mu_{1}$ defined on the Riemann surface $\tilde{\R}_{N}$ on which non-symmetric functions of $N$ Bethe roots live.
\end{subsection}

\begin{subsection}{Large $L$ asymptotics in the KPZ regime}
The KPZ regime of TASEP, reached in the limit $L,N\to\infty$ with fixed density $\rho=N/L$ (or rather $N/L\to\rho$, with $L$ and $N$ co-prime in order to use the expression (\ref{P[H>=U] exp})) and corresponding to an infinite genus limit for $\RN$, is obtained on the time scale $t_{\ell}\sim L^{3/2}$. More precisely, we consider the scalings
\begin{eqnarray}
&& t_{\ell}=\frac{\tau_{\ell}\,L^{3/2}}{\sqrt{\rho(1-\rho)}}\\
&& i_{\ell}=(1-2\rho)t_{\ell}+x_{\ell}L\\
&& H_{i_{\ell}}^{(0)}+U_{\ell}=\rho(1-\rho)t_{\ell}+\mathcal{H}L+\sqrt{\rho(1-\rho)L}\,h_{\ell}\;.
\end{eqnarray}
The initial condition must be chosen so that the local density of particles approaches a regular enough function $\rho(x)$. If $\rho(x)$ is a generic non-constant function, the constant $\mathcal{H}$ is the contribution of the deterministic Burgers' equation on the whole hydrodynamic scale $t\sim L$, in particular $\mathcal{H}=\max(-\rho(1-x_{0}),-(1-\rho)x_{0})$ for domain wall initial condition $x_{k}^{(0)}=k+(x_{0}-\rho)L$ modulo $L$ with $0\leq x_{0}\leq1$, and the statistics of the heights $h_{\ell}$ is expected to be described by the KPZ fixed point with sharp-wedge initial condition and periodic boundaries in that case. If on the other hand $\rho(x)\simeq\rho+\sigma(x)/\sqrt{L}$, which is in particular the case for the stationary initial condition where the height function $h_{0}(x)=\int_{0}^{x}\rmd u\,\sigma(u)$ is a Brownian bridge, one has $\mathcal{H}=0$ and the statistics of the heights $h_{\ell}$ is expected to be described by the KPZ fixed point with initial condition $h_{0}$ and periodic boundaries.

For any sheet $J\subset\[1,L\]$, $|J|=N$, the elements of the set $J-(N+1)/2$ may be interpreted as pseudo-momenta of quasi-particles. The principal sheet $J=\[1,N\]$ corresponds to a filled Fermi sea, with all pseudo-momenta between $-N/2$ and $N/2$ occupied. Only sheets $J$ corresponding to particle-hole excitations close to the edges of the Fermi sea contribute to the KPZ regime of TASEP. Such sheets are parametrized by two finite sets $P,H\subset\Z+1/2$ with $|P|=|H|$, as $J=J_{P,H}$ with
\begin{equation}
\label{J[P,H]}
\fl\hspace{1mm}
J_{P,H}=\Big(\[1,N\]\setminus((1/2-H_{-})\cup(N+1/2-H_{+}))\Big)\cup\Big((N+1/2-P_{-})\cup(L+1/2-P_{+})\Big)
\end{equation}
for large enough $L,N$. The notations $P_{+}$, $H_{+}$ (respectively $P_{-}$, $H_{-}$) refer to the positive (resp. negative) elements of $P$ and $H$. The principal sheet corresponds to both $P$ and $H$ equal to the empty set $\emptyset$.

From (\ref{Aout}), (\ref{Ain}), analytic continuations on $\RN$ from the sheet $J_{P,H}$ crossing the cut finitely many times only lead for large enough $L,N$ to sheets $J_{P',H'}$ with $P'\ominus H'=(P\ominus H)-m$ where $m\in\mathbb{Z}$ and $\ominus$ is the symmetric difference operator $P\ominus H=(P\cup H)\setminus(P\cap H)$. More precisely, describing analytic continuation from $J_{P,H}$ to $J_{P',H'}$ by a sequence of operators $A_{\text{in}}$, $A_{\text{out}}$, $A_{\text{in}}^{-1}$, $A_{\text{out}}^{-1}$, the integer $m$ is equal to the number operators $A_{\text{in}}$, $A_{\text{out}}$ in the sequence, corresponding to analytic continuation from above the cut in figure~\ref{fig a.c.}, minus the number of operators $A_{\text{in}}^{-1}$, $A_{\text{out}}^{-1}$ in the sequence, corresponding to analytic continuation from below the cut in figure~\ref{fig a.c.}. We conclude that the (non-compact) Riemann surface $\R_{\text{KPZ}}$ obtained in the KPZ limit must have infinitely many connected components $\R_{\text{KPZ}}^{\Delta}$ indexed by equivalence classes of sets $\Delta=P\ominus H$ under $\Delta\equiv\Delta+1$.

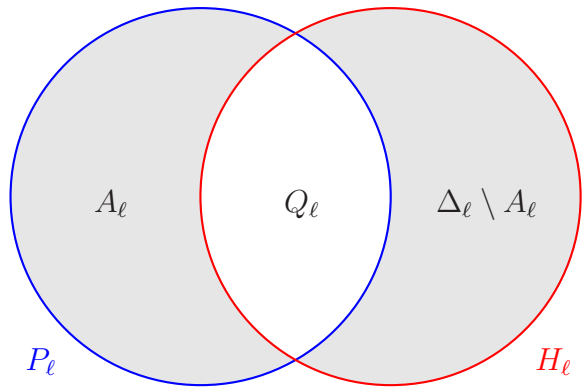
\begin{figure}
	\begin{center}
		\begin{picture}(75,50)
			\put(25,25){\color[rgb]{0.9,0.9,0.9}\circle*{50}}
			\put(50,25){\color[rgb]{0.9,0.9,0.9}\circle*{50}}
			\put(0,0){\color{white}\polygon*(37.5,3.34936)(42.1537,6.81347)(45.2052,10.2776)(47.3215,13.7417)(48.7539,17.2058)(49.6221,20.6699)(49.985,24.134)(49.8646,27.5981)(49.2539,31.0622)(48.1138,34.5263)(46.36,37.9904)(43.8215,41.4545)(40.1079,44.9186)(37.5,46.6506)(32.8463,43.1865)(29.7948,39.7224)(27.6785,36.2583)(26.2461,32.7942)(25.3779,29.3301)(25.015,25.866)(25.1354,22.4019)(25.7461,18.9378)(26.8862,15.4737)(28.64,12.0096)(31.1785,8.54552)(34.8921,5.08142)}
			\put(25,25){\thicklines\color{blue}\circle{50}}
			\put(50,25){\thicklines\color{red}\circle{50}}
			\put(2,2){\color{blue}$P_{\ell}$}
			\put(69,2){\color{red}$H_{\ell}$}
			\put(11,23){$A_{\ell}$}
			\put(36,23){$Q_{\ell}$}
			\put(56,23){$\Delta_{\ell}\setminus A_{\ell}$}
		\end{picture}
	\end{center}
	\caption{Graphical representation of the sets of half-integers appearing in (\ref{P[h>u] KPZ}). The greyed area corresponds to $\Delta_{\ell}=P_{\ell}\ominus H_{\ell}$.}
	\label{fig sets}
\end{figure}

Since the Riemann surface $\RN$ breaks down into connected components $\R_{\text{KPZ}}^{\Delta}$ in the KPZ regime, the starting point $O=[0,\[1,N\]]$ of the integrals in (\ref{P[H>=U] exp}) must be changed before taking the large $L$ limit. For the sheet $J_{\ell}=J_{P_{\ell},H_{\ell}}$ with $\Delta_{\ell}=P_{\ell}\ominus H_{\ell}$ and $Q_{\ell}=P_{\ell}\cap H_{\ell}$, using the results from \ref{appendix integral formulas} and comparing (\ref{P[H>=U] exp}) with (\ref{P[H>=U]}), we choose to replace $O$ by $O_{\Delta_{\ell}}^{A_{\ell}}=[0,J_{A_{\ell},\Delta_{\ell}\setminus A_{\ell}}]$ (i.e. $O_{\Delta_{\ell}}^{A_{\ell}}=O_{A_{\ell},\Delta_{\ell}\setminus A_{\ell}}$ in the notations of \ref{appendix integral formulas}), where $A_{\ell}=P_{\ell}\setminus Q_{\ell}$ and $\Delta_{\ell}\setminus A_{\ell}=H_{\ell}\setminus Q_{\ell}$, see figure~\ref{fig sets}, such that $A_{\ell}$ and $\Delta_{\ell}\setminus A_{\ell}$ have symmetric difference $\Delta_{\ell}$ and an empty intersection. The summation over sets $J_{\ell}$ in (\ref{P[H>=U] exp}) thus reduces in the KPZ regime to sums over finite sets of half-integers $\Delta_{\ell},Q_{\ell}\sqsubset\Z+1/2$ with $\Delta_{\ell}\cap Q_{\ell}=\emptyset$, and $A_{\ell}\subset\Delta_{\ell}$ with the constraint $|A_{\ell}|=|\Delta_{\ell}\setminus A_{\ell}|$ equivalent to $|P_{\ell}|=|H_{\ell}|$.

The Riemann surface $\R_{\text{KPZ}}$ was introduced in \cite{P2020.1} as the natural domain of the functions $\chi_{P,H}(v)$, defined for $-\pi<\Im~v<\pi$ by
\begin{equation}
\label{chiPH}
\fl\hspace{2mm}
\chi_{P,H}(v)=-\frac{\Li_{5/2}(-\rme^{v})}{\sqrt{2\pi}}
+\sum_{a\in P}\frac{(4\rmi\pi a)^{3/2}(1-\frac{v}{2\rmi\pi a})^{3/2}}{3}
+\sum_{a\in H}\frac{(4\rmi\pi a)^{3/2}(1-\frac{v}{2\rmi\pi a})^{3/2}}{3}\;,
\end{equation}
with the usual choice of branch cut $\mathbb{R}^{-}$ for the logarithm and the power $3/2$, and where the polylogarithm $\Li_{s}(z)$, analytic for $z\in\C\setminus[1,\infty)$, is equal for $|z|<1$ to $\Li_{s}(z)=\sum_{k=1}^{\infty}z^{k}/k^{s}$. Writing $\sum_{a\in P}+\sum_{a\in H}=\sum_{a\in P\setminus H}+\sum_{a\in H\setminus P}+2\sum_{a\in P\cap H}=\sum_{a\in P\ominus H}+2\sum_{a\in P\cap H}$ and defining $\Delta=P\ominus H$, $Q=P\cap H$ and $A=P\setminus H$, we observe from (\ref{chiPH}) that the sheets of the Riemann surface, which are labelled by the triplet $(A,\Delta\setminus A,Q)$ in $\RN$, depend in fact only of the pair $(\Delta,Q)$ in $\R_{\text{KPZ}}^{\Delta}$. Labelling the points of $\R_{\text{KPZ}}$ by $[v,(\Delta,Q)]$ with the change of variable $C=\rme^{v}$ from the notations for $\RN$, the points $O_{\Delta}^{A}$ introduced above as starting points of integrals then reduce to the same point $O_{\Delta}=[-\infty,(\Delta,\emptyset)]$ in $\R_{\text{KPZ}}^{\Delta}$.

A straightforward residue calculation, see \ref{appendix large L}, gives the large $L$ asymptotics with fixed $\rho=N/L$ of the functions $\eta$ and $\mu_{1}$ appearing in (\ref{P[H>=U] exp}). We consider two finite sets of half-integers $P$ and $H$ with $|P|=|H|$ and define $\Delta=P\ominus H$, $Q=P\cap H$, $A=P\setminus H$ as above. On the sheet $J_{P,H}$ of $\RN$ defined in (\ref{J[P,H]}), one has for fixed $C\in\C\setminus\mathbb{R}^{-}$
\begin{equation}
\label{eta asymptotics}
\eta([C,J_{P,H}])
=\sqrt{\frac{\rho(1-\rho)}{L}}\,\chi_{P,H}'(\log C)
+\mathcal{O}(L^{-1})
\end{equation}
and
\begin{equation}
\label{mu1 asymptotics}
\mu_{1}([C,J_{P,H}])
=-\sqrt{\frac{1-\rho}{\rho\,L}}\,\chi_{P,H}''(\log C)
+\mathcal{O}(L^{-3/2})\;.
\end{equation}
The convergence is however not uniform near $C=0$ when $(P,H)\neq(\emptyset,\emptyset)$ because fractional powers $C^{1/N}$, $C^{1/(L-N)}$ from (\ref{yj C0}) do not cancel except in the principal sheet, and integrals from $O_{\Delta}^{A}$ contribute additional terms at large $L$. Writing $m=|A|=|\Delta\setminus A|$ and replacing $\eta$ by $\eta+m$ and $\mu_{1}$ by $\mu_{1}+m/N$ in order to ensure convergence of the integrals, one has
\begin{eqnarray}
&&\fl\hspace{5mm} \int_{O_{\Delta}^{A}}^{[C,J_{P,H}]}\!\omega_{\eta+m}\simeq mL+mL\log(\rho^{\rho}(1-\rho)^{1-\rho})+m\log C+2\rmi\pi m\,w(J_{A,\Delta\setminus A}\to J_{P,H})\nonumber\\
&&\fl\hspace{36mm} -2\rmi\pi(1-\rho)\sum_{a\in A}a-2\rmi\pi\rho\sum_{a\in\Delta\setminus A}a+\frac{\sqrt{\rho(1-\rho)}}{\sqrt{L}}\,\chi_{P,H}(\log C)
\end{eqnarray}
with $w(J_{A,\Delta\setminus A}\to J_{P,H})$ the winding number around $0$ of the path from $J_{A,\Delta\setminus A}$ to $J_{P,H}$ in the integral,
\begin{equation}
\int_{O_{\Delta}^{A}}^{[C,J_{P,H}]}\!\omega_{\mu_{1}+m/N}\simeq m\log(\rho(1-\rho)^{\frac{1-\rho}{\rho}})-\sqrt{\frac{1-\rho}{\rho}}\,\frac{\chi_{P,H}'(\log C)}{\sqrt{L}}\;,
\end{equation}
and
\begin{equation}
\rme^{\frac{\rho L}{1-\rho}\int_{O_{\Delta}^{A}}^{[C,J_{P,H}]}\!\omega_{(\mu_{1}+m/N)^{2}}}\simeq
\Big(\frac{\rho^{\frac{2-\rho}{1-\rho}}(1-\rho)^{\frac{1+\rho}{\rho}}L^{2}}{16}\Big)^{m^{2}}
\,\rme^{\rint_{O_{\Delta}}^{[\log C,(\Delta,Q)]}\rmd v\,\chi_{\cdot,\cdot}''(v)^{2}}\;,
\end{equation}
where the differential $\rmd v\,\chi_{\cdot,\cdot}''(v)^{2}$ is understood away from branch points as $\rmd v\,\chi_{\tilde{P},\tilde{H}}''(v)^{2}$ with $\tilde{P}$, $\tilde{H}$ corresponding to the sheets encountered on the path from $O_{\Delta}$ to $[\log C,(\Delta,Q)]$ in $\R_{\text{KPZ}}$, and $\rint$ denoting a natural regularization of the integral by subtracting the divergent terms at $v=-\infty$, see \cite{P2020.1}.

Finally, considering two sheets $J_{P_{1},H_{1}}$, $J_{P_{2},H_{2}}$ of $\RNp$ with symmetric differences $\Delta_{i}=P_{i}\ominus H_{i}$, $i=1,2$, and two initial sheets $J_{A_{1},\Delta_{1}\setminus A_{1}}$, $J_{A_{2},\Delta_{2}\setminus A_{2}}$ with $m_{i}=|A_{i}|=|\Delta_{i}\setminus A_{i}|$, $i=1,2$, one has
\begin{eqnarray}
\label{int gamma asymptotics}
&&\fl\hspace{2mm} \rme^{\frac{\rho L}{1-\rho}\int_{\gamma}\frac{\rmd B}{B}\,\mathcal{A}(\mu_{1}([C_{1}B,\cdot])\mu_{1}([C_{2}B,\cdot]))}\simeq
\Big(\frac{\rho^{\frac{2-\rho}{1-\rho}}(1-\rho)^{\frac{1+\rho}{\rho}}L^{2}}{16}\Big)^{m_{1}m_{2}}
\rme^{\rint_{\beta}\rmd v\,\mathcal{A}(\chi_{\cdot,\cdot}''(v+\log C_{1})\chi_{\cdot,\cdot}''(v+\log C_{2}))}\nonumber\\
&&\hspace{30mm} \times\rme^{\frac{2\rmi\pi}{1-\rho}\big(m_{1}\,w(J_{A_{2},\Delta_{2}\setminus A_{2}}\to J_{P_{2},H_{2}})+m_{2}\,w(J_{A_{1},\Delta_{1}\setminus A_{1}}\to J_{P_{1},H_{1}})\big)}\;,
\end{eqnarray}
where $\gamma\subset\C\setminus\{0,-1\}$ lifts for the pair $([C_{1}B,\cdot],[C_{2}B,\cdot])\in\RNp\times\RNp$ to a path from $(O_{\Delta_{1}}^{A_{1}},O_{\Delta_{2}}^{A_{2}})$ to $([C_{1},J_{P_{1},H_{1}}],[C_{2},J_{P_{2},H_{2}}])$, and $\beta$ is the analogue of $\gamma$ in $\R_{\text{KPZ}}^{\Delta_{1}}\times\R_{\text{KPZ}}^{\Delta_{2}}$ after the change of variable $B=\rme^{v}$.

Using the identities above, we eventually recover \footnote{The function $\rme^{2K^{\Delta_{1},\Delta_{2}}}$ from \cite{P2020.1}, defined there by analytic continuations on $\R_{\text{KPZ}}^{\Delta_{1}}\times\R_{\text{KPZ}}^{\Delta_{2}}$, is in particular equal to $\rme^{\rint_{\beta}\rmd v\,\mathcal{A}(\chi_{\cdot,\cdot}''(v+\log C_{1})\chi_{\cdot,\cdot}''(v+\log C_{2}))}$ with the path $\beta$ as in (\ref{int gamma asymptotics}).} the expressions from \cite{P2020.1} after straightforward calculations. One has
\begin{eqnarray}
\label{P[h>u] KPZ}
&&\fl\hspace{2mm}
\P(h(x_{\ell},\tau_{\ell})> u_{\ell},\ell=1,\ldots,n|h_{0})
=\oint_{|C_{n}|<\ldots<|C_{1}|}
\Big(\prod_{\ell=1}^{n}\frac{\rmd C_{\ell}}{2\rmi\pi\,C_{\ell}}\Big)
\Big(\prod_{\ell=1}^{n}\sum_{\Delta_{\ell}\sqsubset\Z+1/2}\sum_{{Q_{\ell}\sqsubset\Z+1/2}\atop{Q_{\ell}\cap\Delta_{\ell}=\emptyset}}\Big)\nonumber\\
&&\hspace{25mm}
\Xi_{x_{1},\ldots,x_{n}}^{\Delta_{1},\ldots,\Delta_{n}}(C_{1},\ldots,C_{n})\;
\Theta_{h_{0}}([\log C_{1},(\Delta_{1},Q_{1})])\;\\
&&\hspace{8mm} \times\frac{\prod_{\ell=1}^{n}\rme^{\rint_{O_{\Delta_{\ell}}}^{[\log C_{\ell},(\Delta_{\ell},Q_{\ell})]}\rmd v\,\big(\chi_{\cdot,\cdot}''(v)^{2}-(u_{\ell}-u_{\ell-1})\chi_{\cdot,\cdot}''(v)+(\tau_{\ell}-\tau_{\ell-1})\chi_{\cdot,\cdot}'(v)\big)}}{\rme^{\rint_{\beta_{\ell,\ell+1}}\rmd v\,\mathcal{A}(\chi_{\cdot,\cdot}''(v+\log C_{\ell})\chi_{\cdot,\cdot}''(v+\log C_{\ell+1}))}}\;,\nonumber
\end{eqnarray}
with the conventions $\tau_{0}=u_{0}=0$. The summations are over finite subsets $\Delta_{\ell}$ of $\Z+1/2$ labelling the connected components of $\R_{\text{KPZ}}$ and $Q_{\ell}$ labelling the sheets of $\R_{\text{KPZ}}^{\Delta_{\ell}}$. As above, the points of $\R_{\text{KPZ}}^{\Delta}$ are written as $[v,(\Delta,Q)]$, and $O_{\Delta}=[-\infty,(\Delta,\emptyset)]$ belongs to the principal sheet of $\R_{\text{KPZ}}^{\Delta}$. The path $\beta_{\ell,\ell+1}\subset\C\setminus2\rmi\pi(\Z+1/2)$ is such that $([v+\log C_{\ell},(\Delta_{\ell},\cdot)],[v+\log C_{\ell+1},(\Delta_{\ell+1},\cdot)])$ lifts to a path on $\R_{\text{KPZ}}^{\Delta_{\ell}}\times\R_{\text{KPZ}}^{\Delta_{\ell+1}}$ from $(O_{\Delta_{\ell}},O_{\Delta_{\ell+1}})$ to $([\log C_{\ell},(\Delta_{\ell},Q_{\ell})],[\log C_{\ell+1},(\Delta_{\ell+1},Q_{\ell+1})])$. The connected component $(\Delta_{1},\ldots,\Delta_{n})$ of $\R_{\text{KPZ}}^{\Delta_{1}}\times\ldots\times\R_{\text{KPZ}}^{\Delta_{n}}$ is weighted in (\ref{P[h>u] KPZ}) by
\begin{eqnarray}
\label{Xi}
&&\fl\hspace{5mm} \Xi_{x_{1},\ldots,x_{n}}^{\Delta_{1},\ldots,\Delta_{n}}(C_{1},\ldots,C_{n})=\\
&& \Big(\prod_{\ell=1}^{n}\sum_{{A_{\ell}\subset\Delta_{\ell}}\atop{|A_{\ell}|=|\Delta_{\ell}\setminus A_{\ell}|}}\Big)
\prod_{\ell=1}^{n}\Big((\rmi/4)^{|\Delta_{\ell}|}V_{A_{\ell}}^{2}V_{\Delta_{\ell}\setminus A_{\ell}}^{2}\rme^{2\rmi\pi(x_{\ell}-x_{\ell-1})(\sum_{a\in A_{\ell}}a-\sum_{a\in\Delta_{\ell}\setminus A_{\ell}}a)}\Big)\nonumber\\
&&\hspace{13mm} \times\prod_{\ell=1}^{n-1}\frac{(1-C_{\ell+1}/C_{\ell})^{|\Delta_{\ell}|/2}\,(1-C_{\ell}/C_{\ell+1})^{|\Delta_{\ell+1}|/2}}{(1-C_{\ell+1}/C_{\ell})\,V_{A_{\ell},A_{\ell+1}}(C_{\ell},C_{\ell+1})\,V_{\Delta_{\ell}\setminus A_{\ell},\Delta_{\ell+1}\setminus A_{\ell+1}}(C_{\ell},C_{\ell+1})}\;,\nonumber
\end{eqnarray}
where $V_{A}^{2}=\prod_{a<b\in A}(\frac{2\rmi\pi a}{4}-\frac{2\rmi\pi b}{4})^{2}$, $V_{A,B}(C_{1},C_{2})=\prod_{a\in A}\prod_{b\in B}(\frac{2\rmi\pi a-\log C_{1}}{4}-\frac{2\rmi\pi b-\log C_{2}}{4})$, and which is non-zero only if all $|\Delta_{\ell}|$ are even. Finally, the initial condition, specified by the initial height function $h_{0}(x)$, enters through the factor $\Theta_{h_{0}}([\log C_{1},(\Delta_{1},Q_{1})])$, which depends on the variable $C_{1}$ only, as pointed out already in \cite{BL2019.2}. This factor is equal for domain wall initial condition $x_{k}^{(0)}=k+(x_{0}-\rho)L$ to $\Theta_{\text{dw}}([v,(\Delta,Q)])=1$ (with $x_{0}$ in (\ref{Xi}) equal to the coefficient $x_{0}$ defining the shift in $x_{k}^{(0)}$), and for stationary initial condition to $\Theta_{\text{stat}}([v,(\Delta,Q)])=\sqrt{2\pi}\,\rme^{-v}\,\chi_{P,H}'(v)$ with $P,H$ such that $P\ominus H=\Delta$, $P\cap H=Q$, and $x_{0}=0$ in (\ref{Xi}).
\end{subsection}

\end{section}

%%%%%%%%%%%%%%%%%
%%             %%
%%  Section C  %%
%%             %%
%%%%%%%%%%%%%%%%%
\begin{section}{Conclusion}
In this paper, we have studied height fluctuation for TASEP with periodic boundary conditions, with an emphasis on tools from algebraic geometry. Our main result (\ref{P[H>=U] exp}) for the joint probability of the height at multiple times, which is equivalent to earlier expressions of Baik and Liu \cite{BL2019.1}, makes the presence of an underlying compact Riemann surface $\RN$ clear. A relatively straightforward large $L$ asymptotic analysis to the KPZ regime, corresponding to the infinite genus limit $\RN\to\R_{\text{KPZ}}$, then leads directly to analogous expressions involving the non-compact Riemann surface $\R_{\text{KPZ}}$ on which half-integer polylogarithms live, already obtained in \cite{P2020.1} in a much less direct way.

A natural extension of the present work would be to consider instead open TASEP connected to two reservoirs of particles, for which new Bethe equations \cite{CN2018.1} very similar to those of periodic TASEP have been recently discovered by Cramp\'e and Nepomechie, leading to asymptotic expressions \cite{GP2020.1} for the spectrum in the KPZ regime involving a Riemann surface built from infinite sums of Lambert functions. Corresponding eigenfunctions of open TASEP are unfortunately currently missing, but methods from algebraic geometry might be helpful.
\end{section}

\appendix
%%%%%%%%%%%%%%%%%
%%             %%
%%  Section A  %%
%%             %%
%%%%%%%%%%%%%%%%%
\begin{section}{Proof of the identity (\ref{SP psi0 psi0})}
\label{appendix SP psi0 psi0}
In this appendix, we prove the identity (\ref{SP psi0 psi0}). We consider two sets of Bethe roots $y_{j}$ and $w_{j}$ solutions of the Bethe equations (\ref{Bethe equations}), with respective fugacities $\gamma_{y}$ and $\gamma_{w}$ assumed to be distinct. Comparing the definitions (\ref{psi r}), (\ref{psi l}) and (\ref{psi0 r}), (\ref{psi0 l}), one has for arbitrary $\gamma$
\begin{equation}
\langle\psi^{0}_{\vec{w}}|\psi^{0}_{\vec{y}}\rangle=\langle\psi_{\vec{w}}(\gamma)|\psi_{\vec{y}}(\gamma)\rangle\;.
\end{equation}
Setting $\gamma=\gamma_{y}$, the right side becomes the scalar product between an off-shell Bethe vector $\langle\psi_{\vec{w}}(\gamma_{y})|$ for which the $w_{j}$ are \emph{not} solution of the Bethe equations with fugacity $\gamma_{y}$, and an on-shell Bethe vector $|\psi_{\vec{y}}(\gamma)\rangle$ for which the $y_{j}$ are solution of the Bethe equations with fugacity $\gamma_{y}$. The Slavnov determinant (\ref{SP off on}) then leads to
\begin{eqnarray}
&&\fl\hspace{10mm}
\langle\psi^{0}_{\vec{w}}|\psi^{0}_{\vec{y}}\rangle=(-1)^{N}\Big(\prod_{j=1}^{N}\frac{(1-y_{j})^{L+1}}{y_{j}^{N}(1-w_{j})^{L}}\Big)\Bigg(\prod_{j=1}^{N}\prod_{k=1}^{N}(y_{j}-w_{k})\Bigg)\\
&&\hspace{5mm}
\times\det\Bigg(\partial_{y_{i}}\Big(\prod_{k=1}^{N}\frac{1}{1-y_{k}/w_{j}}+\rme^{L\gamma_{y}}(1-w_{j})^{L}\prod_{k=1}^{N}\frac{1}{1-w_{j}/y_{k}}\Big)\Bigg)_{i,j\in\[1,N\]}\;.\nonumber
\end{eqnarray}
Computing explicitly the derivative with respect to $y_{i}$ and using the fact that the $w_{j}$ are solution of the Bethe equations (\ref{Bethe equations}) with fugacity $\gamma_{w}$, we obtain after small simplifications
\begin{eqnarray}
&&
\langle\psi^{0}_{\vec{w}}|\psi^{0}_{\vec{y}}\rangle=\Big(\prod_{j=1}^{N}\frac{(1-y_{j})^{L+1}}{(1-w_{j})^{L}}\Big)\\
&&\hspace{25mm}
\times\det\Bigg(\frac{1}{w_{j}-y_{i}}\prod_{k=1}^{N}\frac{w_{k}}{y_{k}}-\frac{\rme^{L(\gamma_{y}-\gamma_{w})}w_{j}/y_{i}}{w_{j}-y_{i}}\Bigg)_{i,j\in\[1,N\]}\;.\nonumber
\end{eqnarray}
The generalized Cauchy determinant identity
\begin{eqnarray}
&&\fl\hspace{5mm}
\det\Big(\frac{A\,w_{j}-B\,y_{i}}{w_{j}-y_{i}}\Big)_{i,j\in\[1,N\]}=(A-B)^{N-1}\,\Big(A\prod_{j=1}^{N}w_{j}-B\prod_{i=1}^{N}y_{i}\Big)\\
&&\hspace{20mm}
\times(-1)^{\frac{N(N+1)}{2}}\,\frac{(\prod_{i=1}^{N}\prod_{j=i+1}^{N}(y_{i}-y_{j}))(\prod_{i=1}^{N}\prod_{j=i+1}^{N}(w_{i}-w_{j}))}{\prod_{i=1}^{N}\prod_{j=1}^{N}(y_{i}-w_{j})}\;,\nonumber
\end{eqnarray}
which is proved easily from the usual Cauchy determinant identity
\begin{equation}
\fl\hspace{10mm}
\det\Big(\frac{1}{w_{j}-y_{i}}\Big)_{i,j\in\[1,N\]}=\frac{(\prod_{i=1}^{N}\prod_{j=i+1}^{N}(y_{i}-y_{j}))(\prod_{i=1}^{N}\prod_{j=i+1}^{N}(w_{i}-w_{j}))}{(-1)^{\frac{N(N+1)}{2}}\prod_{i=1}^{N}\prod_{j=1}^{N}(y_{i}-w_{j})}
\end{equation}
by writing $\frac{A\,w_{j}-B\,y_{i}}{w_{j}-y_{i}}=A+\frac{(A-B)\,y_{i}}{w_{j}-y_{i}}$ and noting that in the expansion of the determinant only the terms with $(A-B)^{N}$ and $A(A-B)^{N-1}$ contribute, finally leads to (\ref{SP psi0 psi0}).
\end{section}

%%%%%%%%%%%%%%%%
%             %%
%  Section B  %%
%             %%
%%%%%%%%%%%%%%%%
\begin{section}{Abelian integrals starting from an arbitrary sheet}
\label{appendix integral formulas}
In this appendix, we state integral formulas analogue to (\ref{identity sum})-(\ref{identity prod prod 12}) but starting from an arbitrary sheet $J_{P,H}$ of the principal connected component $\RNp$ of $\RN$. The initial sheet $J_{P,H}$ is parametrized as in (\ref{J[P,H]}) in terms of two finite sets $P,H\subset\Z+1/2$, with cardinal $m=|P|=|H|$.

We introduce the point $O_{P,H}=[0,J_{P,H}]\in\RNp$, and the functions $\mu_{1,m}([C,J])=\mu_{1}([C,J])+m/N$ and $\eta_{m}([C,J])=\eta([C,J])+m$, such that the differentials $\omega_{\eta_{m}}$, $\omega_{\mu_{1,m}}$ and $\omega_{\mu_{1,m}^{2}}$ are integrable at the point $O_{P,H}$.

Let $J$ be a sheet of $\RNp$. From the expression (\ref{yj'}) of $y_{j}'(C)$, and the small $C$ behaviour (\ref{yj C0}) of $y_{j}(C)$ to fix the constant of integration, one finds after some calculations the identities
\begin{equation}
\sum_{j\in J}\frac{y_{j}(C)}{1-y_{j}(C)}=-m+\frac{1}{L}\int_{O_{P,H}}^{[C,J]}\omega_{\eta_{m}}-\frac{N}{L}\int_{O_{P,H}}^{[C,J]}\omega_{\mu_{1,m}}\;,
\end{equation}
\begin{eqnarray}
&& \prod_{j\in J}(1-y_{j}(C))=(\rho^{N}(1-\rho)^{L-N}C)^{-\frac{m}{L-N}}
\,\rme^{\frac{2\rmi\pi}{L-N}\!\sum\limits_{a\in P}\!a}\\
&&\hspace{45mm} \times\rme^{-\frac{2\rmi\pi m\,w(J_{P,H}\to J)}{L-N}}
\exp\Big(\frac{N}{L-N}\int_{O_{P,H}}^{[C,J]}\omega_{\mu_{1,m}}\Big)\;,\nonumber
\end{eqnarray}
\begin{eqnarray}
&& \prod_{j\in J}\frac{y_{j}(C)}{\rho^{N}(1-\rho)^{L-N}C}=(\rho^{N}(1-\rho)^{L-N}C)^{-\frac{mL}{N(L-N)}}
\,\rme^{\frac{2\rmi\pi}{L-N}\!\sum\limits_{a\in P}\!a\,+\,\frac{2\rmi\pi}{N}\!\sum\limits_{a\in H}\!a}\\
&&\hspace{45mm} \times\rme^{-\frac{2\rmi\pi Lm\,w(J_{P,H}\to J)}{N(L-N)}}
\exp\Big(\frac{L}{L-N}\int_{O_{P,H}}^{[C,J]}\omega_{\mu_{1,m}}\Big)\;,\nonumber
\end{eqnarray}
and
\begin{eqnarray}
&&\fl \Big(\prod_{j\in J}^{N}\frac{y_{j}(C)(1-y_{j}(C))}{N+(L-N)y_{j}(C)}\Big)
\prod_{{j,k\in J}\atop{j<k}}(y_{j}(C)-y_{k}(C))^{2}\nonumber\\
&&\fl =\frac{(-1)^{\frac{N(N-1)}{2}}\,(-1)^{m}\,(\rho^{N}(1-\rho)^{L-N}C)^{\frac{m^{2}L}{N(L-N)}-\frac{2mL}{L-N}+N}}{(\rho(1-\rho))^{m}L^{2m}}\\
&&\fl\hspace{4mm} \times\rme^{\frac{2\rmi\pi(2N-2m+1)}{L-N}\!\sum\limits_{a\in P}\!a\,-\,\frac{2\rmi\pi}{N}\!\sum\limits_{a\in H}\!a}
\Big(\prod_{a<b\in P}(\rme^{\frac{2\rmi\pi a}{L-N}}-\rme^{\frac{2\rmi\pi b}{L-N}})^{2}\Big)
\Big(\prod_{a<b\in H}(\rme^{-\frac{2\rmi\pi a}{N}}-\rme^{-\frac{2\rmi\pi b}{N}})^{2}\Big)\nonumber\\
&&\fl\hspace{4mm} \times\rme^{2\rmi\pi\,w(J_{P,H}\to J)\,\big(\frac{Lm^{2}}{N(L-N)}-\frac{2Lm}{L-N}\big)}
\exp\Big(\frac{NL}{L-N}\int_{O_{P,H}}^{[C,J]}\omega_{\mu_{1,m}^{2}}+\frac{2(N-m)L}{L-N}\int_{O_{P,H}}^{[C,J]}\omega_{\mu_{1,m}}\Big)\;,\nonumber
\end{eqnarray}
where $w(J_{P,H}\to J)$ is the winding number around $0$ of $\gamma=\pi_{N}(\Gamma)\subset\Ch$, with $\Gamma\subset\RNp$ the path of integration from $[C,J_{P,H}]$ to $[C,J]$ chosen for the integrals. We emphasize that by definition, all these expressions are independent of the initial sheet $J_{P,H}$, but depend only on the final point $[C,J]$.

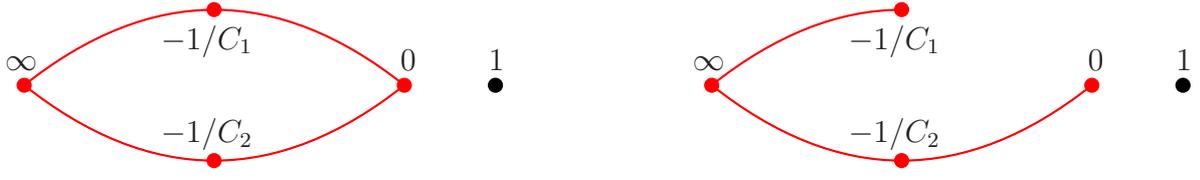
\begin{figure}
	\begin{center}
		\begin{tabular}{ccc}
			\begin{picture}(62,21)
				\put(-2.5,12){$\infty$}
				\put(18,14.8){$-1/C_{1}$}
				\put(18,2.5){$-1/C_{2}$}
				\put(49.5,12){$0$}
				\put(61.1,12){$1$}
				\color{red}\thicklines
				\qbezier(0,10)(12.5,20)(25,20)\qbezier(25,20)(37.5,20)(50,10)
				\qbezier(0,10)(12.5,0)(25,0)\qbezier(25,0)(37.5,0)(50,10)
				\put(0,10){\circle*{2}}
				\put(25,0){\circle*{2}}
				\put(25,20){\circle*{2}}
				\put(50,10){\circle*{2}}
				\color{black}
				\put(62,10){\circle*{2}}
			\end{picture}
			&\hspace*{20mm}&
			\begin{picture}(62,21)
				\put(-2.5,12){$\infty$}
				\put(18,14.8){$-1/C_{1}$}
				\put(18,2.5){$-1/C_{2}$}
				\put(49.5,12){$0$}
				\put(61.1,12){$1$}
				\color{red}\thicklines
				\qbezier(0,10)(12.5,20)(25,20)
				\qbezier(0,10)(12.5,0)(25,0)\qbezier(25,0)(37.5,0)(50,10)
				\put(0,10){\circle*{2}}
				\put(25,0){\circle*{2}}
				\put(25,20){\circle*{2}}
				\put(50,10){\circle*{2}}
				\color{black}
				\put(62,10){\circle*{2}}
			\end{picture}
		\end{tabular}
	\end{center}
	\caption{Branch cuts of the function $B\mapsto\mu_{1}([C_{1}B,J_{1}])\mu_{1}([C_{2}B,J_{2}])$ with $J_{1},J_{2}\neq\[1,N\]$ (left) and $J_{1}=\[1,N\]$, $J_{2}\neq\[1,N\]$ (right).}
	\label{fig cuts D2}
\end{figure}

Similarly, considering two sheets $J_{1},J_{2}$ of $\RNp$ and defining initial sheets $J_{P_{1},H_{1}}$, $J_{P_{2},H_{2}}$ with $m_{1}=|P_{1}|=|H_{1}|$, $m_{2}=|P_{2}|=|H_{2}|$ in $\RNp$, one has
\begin{eqnarray}
&&\fl \prod_{j_{1}\in J_{1}}\prod_{j_{2}\in J_{2}}(y_{j_{1}}(C_{1})-y_{j_{2}}(C_{2}))
=(-1)^{m_{1}+m_{1}m_{2}}
(\rho^{N}(1-\rho)^{L-N}(C_{1}-C_{2}))^{N-m_{1}-m_{2}}\nonumber\\
&&\fl\hspace{7mm} \times(\rho^{N}(1-\rho)^{L-N})^{\frac{m_{1}m_{2}(L-2N)}{N(L-N)}}(\rho^{N}(1-\rho)^{L-N}C_{1})^{-\frac{m_{1}(N-m_{2})}{L-N}}(\rho^{N}(1-\rho)^{L-N}C_{2})^{-\frac{m_{2}(N-m_{1})}{L-N}}\nonumber\\
&&\fl\hspace{7mm} \times\Big(\prod_{a\in P_{1}}\prod_{b\in P_{2}}(\rme^{\frac{2\rmi\pi a}{L-N}}C_{1}^{-\frac{1}{L-N}}-\rme^{\frac{2\rmi\pi b}{L-N}}C_{2}^{-\frac{1}{L-N}})\Big)
\Big(\prod_{a\in H_{1}}\prod_{b\in H_{2}}(\rme^{-\frac{2\rmi\pi a}{N}}C_{1}^{\frac{1}{N}}-\rme^{-\frac{2\rmi\pi b}{N}}C_{2}^{\frac{1}{N}})\Big)\nonumber\\
&&\fl\hspace{7mm} \times\rme^{\frac{2\rmi\pi(N-m_{2})}{L-N}\!\sum\limits_{a\in P_{1}}\!a\,+\,\frac{2\rmi\pi(N-m_{1})}{L-N}\!\sum\limits_{a\in P_{2}}\!a}
\;\rme^{2\rmi\pi w(\gamma)\big(\frac{Lm_{1}m_{2}}{N(L-N)}-\frac{L(m_{1}+m_{2})}{L-N}\big)}\\
&&\fl\hspace{7mm}
\times\exp\Bigg(
\frac{NL}{L-N}\int_{\gamma}\frac{\rmd B}{B}\,\mathcal{A}(\mu_{1,m_{1}}([C_{1}B,\cdot])\mu_{1,m_{2}}([C_{2}B,\cdot]))\nonumber\\
&&\hspace{20mm} +\frac{(N-m_{2})L}{L-N}\int_{O_{P_{1},H_{1}}}^{[C_{1},J_{1}]}\omega_{\mu_{1,m_{1}}}
+\frac{(N-m_{1})L}{L-N}\int_{O_{P_{2},H_{2}}}^{[C_{2},J_{2}]}\omega_{\mu_{1,m_{2}}}
\Bigg)\;.\nonumber
\end{eqnarray}
where $\gamma$ is a path from $0$ to $1$ in $\C$, $w(\gamma)$ its winding number around $0$, and $\mathcal{A}(\ldots)$ meaning analytic continuation of the couple $([C_{1}B,\cdot],[C_{2}B,\cdot])$ on the path $\Gamma:(O_{P_{1},H_{1}},O_{P_{2},H_{2}})\to([C_{1},J_{1}],[C_{2},J_{2}])$ obtained by lifting $\gamma$ to $\RNp\times\RNp$.

It is always possible to find such a path $\Gamma$ in $\RNp\times\RNp$. Indeed, by definition, one can always find a path from $(O_{P_{1},H_{1}},O_{P_{2},H_{2}})$ to $([C,\[1,N\]],[C,K])$ for some $K\subset\[1,L\]$ by crossing the cuts $(0,-1/C_{1})$, $(0,-1/C_{2})$, $(-1/C_{1},\infty)$, $(-1/C_{2},\infty)$ for $B$. Then, in the sheet $(\[1,N\],K)$ of $\RNp\times\RNp$ the branch cut $(0,-1/C_{1})$ disappears, see figure~\ref{fig cuts D2}, and one can then find a path from the sheet $(\[1,N\],K)$ to the sheet $(\[1,N\],\[1,N\])$ by crossing the cuts $(0,-1/C_{2})$ and $(-1/C_{2},\infty)$ for $B$. In other words, the fibre product $\RNp*\RNp$ generated by analytic continuations in $B$ of couples $([C_{1}B,J_{1}],[C_{2}B,J_{2}])\in\RNp\times\RNp$ is a connected space.
\end{section}

%%%%%%%%%%%%%%%%
%             %%
%  Section C  %%
%             %%
%%%%%%%%%%%%%%%%
\begin{section}{Large \texorpdfstring{$L$}{L} asymptotics}
\label{appendix large L}
We consider the meromorphic function $\mu_{1}$ on $\RN$ defined in (\ref{mu1}). Since $y_{j}(C)\to0$ when $1\leq j\leq N$ and $y_{j}(C)\to\infty$ when $N+1\leq j\leq L$, the $C\to0$ expansion of $\mu_{1}([C,J])$ in the principal sheet $J=\[1,N\]$ can be computed by residues as in \cite{DL1998.1}, using
\begin{equation}
\sum_{j=1}^{N}\frac{1}{N+(L-N)y_{j}(C)}=\oint_{\gamma_{0}}\frac{\rmd y}{2\rmi\pi}\,\frac{\partial_{y}\log P(y,C)}{N+(L-N)y}\;,
\end{equation}
with $P$ the polynomial defined in (\ref{P}) and $\gamma_{0}$ a small contour encircling $0$ once in the positive direction. Expanding in powers of $C$ and computing residues explicitly, one has
\begin{eqnarray}
&& -1+\sum_{j=1}^{N}\frac{1}{N+(L-N)y_{j}(C)}\\
&& =\sum_{k=1}^{\infty}\frac{(-1)^{k}(\rho^{N}(1-\rho^{L-N})C)^{k}}{kN}
\sum_{m=0}^{\infty}m\Big(\frac{1-\rho}{\rho}\Big)^{m}{{kL}\choose{kN-m}}\nonumber\;.
\end{eqnarray}
At large $L$ with fixed $\rho=N/L$, the sum over $m$ is dominated by the regime $m\sim\sqrt{L}$. Writing the asymptotic expansion of the binomial coefficient, the Euler-Maclaurin formula finally gives (\ref{mu1 asymptotics}) with $P=H=\emptyset$.

The contribution of particle-hole excitations for a sheet $J=J_{P,H}$ defined in (\ref{J[P,H]}) can be computed by writing
\begin{eqnarray}
&& \sum_{j\in J_{P,H}}\frac{1}{N+(L-N)y_{j}(C)}=
\sum_{j=1}^{N}\frac{1}{N+(L-N)y_{j}(C)}\\
&&\hspace{7mm} -\sum_{a\in H_{-}}\frac{1}{N+(L-N)y_{1/2-a}(C)}
-\sum_{a\in H_{+}}\frac{1}{N+(L-N)y_{N+1/2-a}(C)}\nonumber\\
&&\hspace{7mm} +\sum_{a\in P_{-}}\frac{1}{N+(L-N)y_{N+1/2-a}(C)}
+\sum_{a\in P_{+}}\frac{1}{N+(L-N)y_{L+1/2-a}(C)}\;,\nonumber
\end{eqnarray}
and using the asymptotics
\begin{eqnarray}
&& y_{N+1/2-a}(C)\simeq-\frac{\rho}{1-\rho}\Bigg(1-\sign~a\;\frac{\sqrt{4\rmi\pi a}\sqrt{1-\frac{\log C}{2\rmi\pi a}}}{\sqrt{\rho(1-\rho)L}}\Bigg)\\
&&\hspace{5mm} y_{1/2-a}(C)\simeq-\frac{\rho}{1-\rho}\Bigg(1+\sign~a\;\frac{\sqrt{4\rmi\pi a}\sqrt{1-\frac{\log C}{2\rmi\pi a}}}{\sqrt{\rho(1-\rho)L}}\Bigg)
\end{eqnarray}
with the convention $y_{j}(C)=y_{j+L}(C)$ finally gives (\ref{mu1 asymptotics}) for general $P,H$ with $|P|=|H|$. The large $L$ asymptotics (\ref{eta asymptotics}) of the function $\eta$ on $\RN$ defined in (\ref{eta}) can be obtained in a similar way.
\end{section}

\vspace{10mm}
%\bibliographystyle{unsrt}
%\bibliography{/users/prolhac/bib/references.bib}
%\bibliography{F:/Donnees/Recherche/bib/references}

\end{document}